\documentclass[a4paper,10pt]{article}
\usepackage{latexsym,amssymb,amsmath,amsthm,amsxtra,xypic}
\usepackage{stmaryrd}
\usepackage{amsfonts}
\usepackage{bbm}
\usepackage{amscd}
\usepackage{mathrsfs}
\usepackage[dvips]{graphicx}
\usepackage[all]{xy}
\usepackage{ulem}

\newcommand{\allowpagebreak}
\allowdisplaybreaks

\newtheorem{thm}{Theorem}[section]
\newtheorem{lem}[thm]{Lemma}
\newtheorem{cor}[thm]{Corollary}
\newtheorem{pro}[thm]{Proposition}
\newtheorem{ex}[thm]{Example}

\theoremstyle{definition}
\newtheorem{defi}[thm]{Definition}
\newtheorem{rmk}[thm]{Remark}

\newcommand {\emptycomment}[1]{}

\newcommand{\HH}{\mathbf H}

\newcommand{\str}{\mathbf {string}}

\newcommand{\Lie}{\mathbf {Lie2}}

\newcommand{\pf}{\noindent{\bf Proof.}\ }

\def\R{\mathbb{R}}

\newcommand{\huaV}{\mathcal{V}}

\newcommand{\gl}{\mathfrak {gl}}

\newcommand{\s}{\mathfrak s}
\newcommand{\li}{\mathfrak l}
\newcommand{\g}{\mathfrak g}

\newcommand{\hg}{\hat{\mathfrak g}}
\newcommand{\tg}{\tilde{\mathfrak g}}
\newcommand{\h}{\mathfrak h}
\newcommand{\frkg}{\mathfrak g}
\newcommand{\frkh}{\mathfrak h}

\newcommand{\gh}{\mathfrak g \oplus \mathfrak h}

\newcommand{\ulam}{^\lambda}
\newcommand{\dlam}{_\lambda}

\newcommand{\dM}{\mathrm{d}}
\newcommand{\E}{\mathrm{E}}

\newcommand{\Hom}{\mathrm{Hom}}

\newcommand{\Cen}{\mathrm{Cen}}
\newcommand{\Ker}{\mathrm{Ker}}

\newcommand{\End}{\mathrm{End}}

\newcommand{\ad}{\mathrm{ad}}

\newcommand{\id}{\mathrm{id}}

\newcommand{\brh}[1]{   [    #1   ]_{\gh}   }
\newcommand{\brt}[1]{   [    #1   ]'   }

\newcommand{\V}{\mathbb{V}}

\begin{document}

\title{ {Deformations of Lie 2-algebras %cohomology and applications
\thanks
 {
Research partially supported by NSFC  (11101179) and SRFDP
(20100061120096).
 }
 \author{\vspace{2mm}Zhangju Liu$^1$, Yunhe Sheng$^2$ and  Tao Zhang$^1$\\
 $^1$Department of Mathematics and LMAM, Peking University,\\
Beijing 100871, China\\
$^2$ Department of Mathematics, Jilin University,
 \\Changchun 130012, Jilin, China\\
Email: liuzj@pku.edu.cn, shengyh@jlu.edu.cn, zhangtao@pku.edu.cn
}
 }}

\date{}
\maketitle



\begin{abstract}
  In this paper, we consider deformations  of Lie $2$-algebras via the cohomology theory. We prove that
  a 1-parameter infinitesimal deformation of a Lie $2$-algebra $\g$  corresponds to a 2-cocycle of $\g$ with the coefficients in the adjoint representation.
   The Nijenhuis operator  for Lie 2-algebras is introduced to describe trivial
  deformations. We also study  abelian extensions of Lie 2-algebras from the viewpoint of  deformations of  semidirect product Lie 2-algebras.
\end{abstract}

\section{Introduction}
Recently, people have paid more attention to higher categorical
structures with motivations from string theory. One way to provide
higher categorical structures is by categorifying existing
mathematical concepts. One of the simplest higher structure is a
$2$-vector space, which is a categorified  vector space.
Furthermore, the notion of a Lie $2$-algebra  is obtained by adding
some weak Lie algebra structures on a $2$-vector space  \cite{Baez},
where the Jacobi identity is replaced by a natural transformation,
called the Jacobiator, with some coherence laws of its own.

For  a Lie algebra $(\li,[\cdot,\cdot]_\li)$, a Nijenhuis operator is
a linear map $N:\li\longrightarrow\li$ satisfying
$$
[Nx, Ny]_\li = N([Nx, y]_\li +[x, Ny]_\li - N[x,y]_\li),
$$
which gives a trivial deformation of  Lie algebra $\li$ and plays
important role in the study of integrability of Hamilton equations
\cite{D, Kosmann3}. In general, a 1-parameter infinitesimal
deformation is controlled  by a 2-cocycle
$\omega:\wedge^2\li\longrightarrow\li$ (See \cite{NR} for more
details). In \cite{grabowski}, the authors identified the role that
Nijenhuis operators  play in the theory of contractions and
deformations of both Lie algebras and Leibniz (Loday) algebras.
Nijenhuis operators for algebras other than Lie algebras, including
for Courant algebroids, can be also found in \cite{AGC, Semisimple
N,  NijenhuisforA, G, Kosmann1, MX}.

In this paper, we study deformations of Lie 2-algebras. To do that,
we give precise formulas for the cohmologies
 of Lie 2-algebras. In the strict case, they are already given in \cite{Sheng1}. A 1-parameter infinitesimal deformation of a Lie 2-algebra  $(\g;\dM_\g,[\cdot,\cdot]_\g,l_3^\g)$ is given by a 2-cocycle $(\omega_1,\omega_2^0,\omega_2^1,\omega_3)$ satisfying that itself defines a Lie 2-algebra structure. We pay special attention to trivial deformations, and introduce the notion of a Nijenhuis operator for Lie 2-algebras. Due to the
abundant content of the corresponding cohomology theory, a Nijenhuis
operator $N$ for a Lie 2-algebra contains $N_0:\g_0\longrightarrow
\g_0$ and $ ~N_1:\g_{-1}\longrightarrow\g_{-1}$ such that some
compatibility conditions are satisfied.

The paper is organized as follows. In Section 2, we recall some
notions of Lie 2-algebras and their  homomorphisms.  We study the
cohomology of Lie 2-algebras in detail, and give precise
 formulas for the coboundary operator. In Section 3, we study the 1-parameter infinitesimal deformation of a
 Lie $2$-algebra $\g$, and give the conditions on generating a 1-parameter infinitesimal deformation as a cocycle
 condition (Theorem \ref{thm:deformation}). We introduce a notion of a Nijenhuis operator,
 which could give trivial deformations (Theorem \ref{thm:Nijenhuis}). We construct examples of Nijenhuis operators for the Lie 2-algebra of string type in term of $\mathcal O$-operators. In Section 4, we study abelian extensions of Lie
  2-algebras. We prove that any abelian extension corresponds to a representation and a 2-cocycle. Thus, an abelian
  extension can be viewed as a deformation of a semidirect product Lie 2-algebra  (Remark \ref{rmk:deform}) and can be
classified by  the second cohomology.

\vspace{3mm}

{\bf Notations:} $\g=\g_0\oplus \g_{-1}$ is a graded vector space, $x,~y,~z,~t,~x_i$ are elements in $\g_0$ and $a,~b,~a_i$ are elements in $\g_{-1}$. $Sym(\V)$ is the symmetric algebra of a graded vector space $\V$, and $\odot V$ is the symmetric algebra of a vector space $V$.

\section{Lie 2-algebras and their cohomologies}\label{sec:cohom}

The category of Lie 2-algebras and the category of 2-term
$L_\infty$-algebras are equivalent. What we call a Lie 2-algebra  is
actually a 2-term $L_\infty$-algebra, see \cite{Baez,LadaMarkl}.

\begin{defi} {\rm\cite{Baez}}\label{defi:Lie 2}
A Lie 2-algebra structure on a graded vector space $\frkg=\g_0\oplus\g_{-1}$ contains of linear maps $\dM_\g: \frkg_{-1}
{\longrightarrow} \frkg_{0}$, $[\cdot,\cdot]_\g:\g_i\wedge\g_j\longrightarrow\g_{i+j}~ (-1\leq i+j\leq0),$
and $l_3^\g:\wedge^3\g_0\longrightarrow\g_{-1}$,
such that the following equalities are satisfied:
\begin{itemize}
\item[\rm(i)] $\dM_\g [x,a]_\g=[x,\dM_\g a]_\g,$
\item[\rm(ii)]  $[\dM_\g a,b]_\g=[a,\dM_\g b]_\g,$
\item[\rm(iii)]  $[[x,y]_\g,z]_\g+c.p.+\dM_\g l_3^\g(x,y,z)=0,$
\item[\rm(iv)]  $[[x,y]_\g,a]_\g+[[y,a]_\g,x]_\g+[[a,x]_\g,y]_\g+l_3^\g(x,y,\dM_\g a)=0,$
\item[\rm(v)]  $l_3^\g([x,y]_\g,z,t)+c.p.=[l_3^\g(x,y,z),t]_\g+c.p.,$
\end{itemize}
where $c.p.$ means cyclic permutation. We denote a Lie $2$-algebra
by $(\g;\dM_\g,[\cdot,\cdot]_\g,l_3^\g)$.
\end{defi}

A Lie 2-algebra is called  {\bf strict} if $l_3=0$ or {\bf skeletal
} if $\dM = 0$.
 Usually, $l_3$ is called the {\bf Jacobiator}, and condition (v) is called the Jacobiator identity. We will denote
$$
J_{l_3^\g}(x,y,z,t)=[l_3^\g(x,y,z),t]_\g+c.p.-(l_3^\g([x,y]_\g,z,t)+c.p.).
$$
Thus, the Jacobiator identity reads $J_{l_3^\g}(x,y,z,t)=0$.

\begin{defi}\label{defi:Lie-2hom}
Let $(\frkg;\dM_\g,[\cdot,\cdot]_\g,l_3^\g)$ and $(\frkg';\dM',[\cdot,\cdot]',l_3')$ be Lie 2-algebras. A
Lie 2-algebra homomorphism $F$ from $\frkg$ to $ \frkg'$ consists of:
 linear maps $F_0:\frkg_0\rightarrow \frkg_0',~F_1:\frkg_{-1}\rightarrow \frkg_{-1}'$
 and $F_{2}: \frkg_{0}\wedge \frkg_0\rightarrow \frkg_{-1}'$,
such that the following equalities hold for all $ x,y,z\in \frkg_{0},
a\in \frkg_{-1},$
\begin{itemize}
\item [$\rm(i)$] $F_0\dM_\g=\dM'F_1$,
\item[$\rm(ii)$] $F_{0}[x,y]_\g-[F_{0}(x),F_{0}(y)]'=\dM'F_{2}(x,y),$
\item[$\rm(iii)$] $F_{1}[x,a]_\g-[F_{0}(x),F_{1}(a)]'=F_{2}(x,\dM_\g a)$,
\item[$\rm(iv)$]
$F_2([x,y]_\g,z)+c.p.+F_1(l_3^\g(x,y,z))=[F_0(x),F_2(y,z)]'+c.p.+l_3'(F_0(x),F_0(y),F_0(z))$.
\end{itemize}\end{defi}
 If $F_2=0$, the homomorphism $F$
is called a strict homomorphism. The identity homomorphism $1_\frkg:
\frkg\to \frkg$ has the identity chain map together with
$(1_\frkg)_2=0$.

Let $F:\frkg\to \frkg'$ and $G:\frkg'\to \frkg''$ be two homomorphisms, then their composition $GF:\frkg\to \frkg''$
is a homomorphism defined as $(GF)_0=G_0\circ F_0:\frkg_0\to \frkg''_0$, $(GF)_1=G_1\circ F_1:\frkg_{-1}\to \frkg''_{-1}$
and
$$(GF)_2=G_2\circ (F_0\times F_0)+G_1\circ F_2:\frkg_0\times \frkg_0\to \frkg''_{-1}.$$
A homomorphism $F:\frkg\to \frkg'$ is called an isomorphism if there exists a homomorphism $F^{-1}:\frkg'\to \frkg$ such that
their composition $F^{-1}F:\frkg\to \frkg$ and $FF^{-1}:\frkg'\to \frkg'$ are all identity homomorphisms. It is easy to show that

\begin{lem}
  Let $F=(F_0,F_1,F_2):\frkg\to \frkg'$ be
a homomorphism. If $F_0, F_1$ are invertible, then $F$ is an isomorphism, and $F^{-1}$ is given by
 $$F^{-1}=(F^{-1}_0, F^{-1}_1, -F^{-1}_1 F_2(F^{-1}_0\times F^{-1}_0)).$$
\end{lem}

In fact, let $(\g';\dM',[\cdot,\cdot]',l_3')$ be a Lie
2-algebra, then one can construct a Lie 2-algebra structure on a graded
vector space $\frkg=\g_0\oplus\g_{-1}$  by use of a triple
$F=(F_0,F_1,F_2):\g\longrightarrow \g'$ with invertible $F_0$ and  $F_1$  as follows, for all $
x,y,z\in \frkg_{0}$ and $a\in \frkg_{-1},$ define
\begin{itemize}
\item [$\rm(i)$] $\dM_\g \triangleq F^{-1}_0\dM'F_1$;
\item[$\rm(ii)$] $[x,y]_\g \triangleq F^{-1}_0 ([F_{0}(x),F_{0}(y)]' +\dM'F_{2}(x,y));$
\item[$\rm(iii)$] $[x,a]_\g \triangleq F^{-1}_1 ([F_{0}(x),F_{1}(a)]'+ F_{2}(x,\dM_\g
a))$;
\item[$\rm(iv)$]
$l_3^\g(x,y,z)\triangleq F^{-1}_1 \{[F_0(x),F_2(y,z)]'-
  F_2([x,y]_\g,z)+c.p. +l_3'(F_0(x),F_0(y),F_0(z))$\}.
\end{itemize}

\begin{pro}\label{pro:obtainLie2}With the above notations, for
 a graded vector space  $\frkg$ with  linear maps
$F_0:\g_0\to \frkg_0'$,
   $F_1:\frkg_{-1}\to \frkg_{-1}'$ and $F_2:\wedge^2\g_0\longrightarrow
   \g_{-1}'$, such that $F_0$ and $F_1$ are invertible, then $(\g,\dM_\g,[\cdot,\cdot]_\g,l_3^\g)$ defined above is a Lie
   $2$-algebra such that  $F: \frkg \rightarrow\frkg' $ is a Lie
   $2$-algebra isomorphism.
\end{pro}
 \pf Since $\g'$ is a
Lie 2-algebra, we have
$$\dM'[F_0x,F_1a]'=[F_0x,F_0\dM_\g a]'.$$
Consider the left hand side, we have
$$\dM'[F_0x,F_1a]'=\dM' F_1[x,a]_\g-\dM' F_2(x,\dM_\g a)=F_0\dM_\g[x,a]_\g-\dM' F_2(x,\dM_\g a).$$
The right hand side is equal to
$F_0[x,\dM_\g a]_\g-\dM' F_2(x,\dM_\g a).$
Thus, we have
\begin{equation}\label{eq:te1}\dM_\g[x,a]_\g=[x,\dM_\g a]_\g,\end{equation}
since $F_0$ is invertible. Similarly, by $[\dM'
F_1a,F_1b]'=[F_1a,\dM' F_1b]'$,  we obtain
\begin{equation}\label{eq:te2}[\dM_\g a,b]_\g=[a,\dM_\g b]_\g.\end{equation}
Furthermore, we have
\begin{eqnarray*}0&=&[[F_0x,F_0y]',F_0z]'+c.p.+\dM' l_3'(F_0x,F_0y,F_0z)\\
&=&[F_0[x,y]_\g,F_0z]'-[\dM' F_2(x,y),F_0z]'+c.p.+\dM' l_3'(F_0x,F_0y,F_0z)\\
&=&F_0[[x,y]_\g,z]_\g-\dM' F_2([x,y]_\g,z)-\dM' [F_2(x,y),F_0z]'+c.p.+\dM' l'_3(F_0x,F_0y,F_0z)\\
&=&F_0[[x,y]_\g,z]_\g+c.p.+\dM' F_1l_3^\g(x,y,z)\\
&=&F_0\big([[x,y]_\g,z]_\g+c.p.+\dM_\g l_3^\g(x,y,z)\big).
\end{eqnarray*}
Thus, we have
\begin{equation}\label{eq:te3}[[x,y]_\g,z]_\g+c.p.+\dM_\g l_3^\g(x,y,z)=0.\end{equation}

Similarly, by $$0=[[F_0x,F_0y]',F_1a]'+[[F_0y,F_1a]',F_0x]'+[[F_1a,F_0x]',F_0y]' +l_3'(F_0x,F_0y,\dM' F_1a),$$  we have
\begin{equation}\label{eq:te4}[[x,y]_\g,a]_\g+[[y,a]_\g,x]_\g+[[a,x]_\g,y]_\g+ l_3^\g(x,y,\dM_\g a)=0.\end{equation}

Finally, by the Jacobiator identity
$l'_3([F_0x,F_0y]',F_0z,F_0t)+c.p.=[l'_3(F_0x,F_0y,F_0z),F_0t]'+c.p.$,
we have
\begin{equation}\label{eq:te5}
  l_3^\g([x,y]_\g,z,t)+c.p.=[l_3^\g(x,y,z),t]_\g+c.p..
\end{equation}
By \eqref{eq:te1}-\eqref{eq:te5}, we deduce that $(\g,\dM_\g,[\cdot,\cdot]_\g,l_3^\g)$ is a Lie 2-algebra.
\qed\vspace{3mm}

The notion of  an $L_\infty$-module was introduced in
\cite{LadaMarkl}. Given a $k$-term complex of vector spaces $\huaV:
V_{-k+1}\stackrel{\partial}{\longrightarrow}\cdots
V_{-1}\stackrel{\partial}{\longrightarrow}V_0$, the endomorphisms
form a strict Lie 2-algebra $\gl(\huaV)$ with the graded commutator
bracket and a differential inherited from $\partial$. This plays the
same role as $\gl(V)$ in the classical case for a vector space $V$
(see \cite{LadaMarkl, shengzhu2}). We say that $\huaV$ is an {\bf
$L_\infty$-module} of an  $L_\infty$-algebra $L$ if there is an
$L_\infty$-morphism $L \to \gl(\huaV)$, in which $\gl(\huaV)$ is
considered as an $L_\infty$-algebra. We can also describe an
$L_\infty$-module by a generalized Chevalley-Eilenberg complex of
$L$. An $L_\infty$-module structure on a graded vector space $\huaV$
is equivalent to a degree 1 differential $D$ on the graded vector
space
$$
 (Sym (L^*[-1])
\otimes \huaV)_n = \oplus_k Sym(L^*[-1])_k \otimes V_{n-k},
 $$
such that $D^2=0.$

In the following, we focus on the 2-term case.  Let $ \mathbb
V:V_{-1}\stackrel{\partial}{\longrightarrow} V_0$ be a 2-term
complex of vector spaces, and we can form a new 2-term complex of
vector spaces $\End(\mathbb V):\End^1(\mathbb
V)\stackrel{\delta}{\longrightarrow} \End^0_\partial(\mathbb V)$ by
defining $\delta(A)=\partial\circ A+A\circ\partial$ for all
$A\in\End^1(\mathbb V)$, where $\End^1(\mathbb V)=\End(V_0,V_{-1})$
and
$$\End^0_\partial(\mathbb V)=\{X=(X_0,X_1)\in \End(V_0,V_0)\oplus \End(V_{-1},V_{-1})|~X_0\circ \partial=\partial\circ X_1\}.$$
Define  $l_2:\wedge^2 \End(\mathbb V)\longrightarrow \End(\mathbb
V)$ by setting:
\begin{equation}
\left\{\begin{array}{l}l_2(X,Y)=[X,Y]_C,\\
l_2(X,A)=[X,A]_C,\\
l_2(A,A')=0,\end{array}\right.\nonumber
\end{equation}
 for all $X,Y\in
\End^0_\partial(\mathbb V)$ and $A,A'\in \End^1(\mathbb V),$ where
$[\cdot,\cdot]_C$ is the graded commutator.

\begin{thm}\label{thm:End(V)} {\rm\cite{LadaMarkl,shengzhu1}} With the above notations,
$(\End(\mathbb V),\delta,l_2)$ is a strict Lie $2$-algebra.
\end{thm}

A representation of a Lie 2-algebra $(\frkg; \dM_\g,[\cdot,\cdot]_\g,l_3^\g)$ on a 2-term complex
 $\mathbb V$ is a Lie 2-algebra homomorphism $\mu=(\mu_0,\mu_1,\mu_2):\frkg\longrightarrow \End(\mathbb V).$ Given
 a representation, the corresponding generalized
Chevalley-Eilenberg complex is given by
{\footnotesize
\begin{equation} \label{eq:complex}
\begin{split}
 & V_{-1}\stackrel{D}{\longrightarrow}\\
 & V_{0}\oplus \Hom(\frkg_{0},V_{-1})\stackrel{D}{\longrightarrow}\\
  &\Hom(\frkg_{0},V_{0})\oplus\Hom(\frkg_{-1},V_{-1})\oplus\Hom(\wedge^2\frkg_0,V_{-1})\stackrel{D}{\longrightarrow}\\
  & \Hom(\frkg_{-1},V_{0})\oplus \Hom(\wedge^2\frkg_0,V_{0})\oplus \Hom(\frkg_0\wedge \frkg_{-1},V_{-1}) \oplus \Hom(\wedge^3\frkg_0,V_{-1})\stackrel{D}{\longrightarrow}\\
  &\Hom(\frkg_0\wedge \frkg_{-1},V_{0})\oplus\Hom(\odot^2\frkg_{-1},V_{-1})\oplus\Hom(\wedge^3\frkg_0,V_{0}) \oplus \Hom(\wedge^2\frkg_{0}\wedge\frkg_{-1},V_{-1})\oplus  \Hom(\wedge^4\frkg_0,V_{-1})\\
  & \stackrel{D}{\longrightarrow}\cdots,
\end{split}
\end{equation}}
where the first line is of degree $-1$, the second line is of degree
0, the third line is of degree 1, and etc. In fact, the degree of an
element in $\Hom((\wedge^k\g_0)\wedge\odot^l\g_{-1},V_s)$ is
$k+2l+s$. The corresponding cohomology is denoted by $\HH(\g;\mu)$
and the differential $D$  can be written  in components:
\begin{equation}
  D=\hat{\dM_\g}+\hat{\partial}+d_\mu+d_{\mu_2}+d_{l_3},
\end{equation}
where each term is defined respectively as follows:

$\bullet$ \, ~~ $\hat{\dM_\g}:\Hom((\wedge^p\frkg_0)\wedge (\odot
^q\frkg_{-1}),V_s)\xrightarrow{~}\Hom((\wedge  ^{p-1}\frkg_0)\wedge
(\odot  ^{q+1}\frkg_{-1}),V_s)$ defined  by
\begin{eqnarray*}
 && \hat{\dM_\g}(f)(x_1,\cdots,x_{p-1},a_1,a_2,\cdots,a_{q+1})\\
 &=&(-1)^{p}\big(f(x_1,\cdots,x_{p-1},\dM_\g
  a_1,a_2,\cdots,a_{q+1})+c.p.(a_1,\cdots,a_{q+1})\big);
\end{eqnarray*}

$\bullet$\,  $\hat{\partial}:\Hom((\wedge  ^p\frkg_0)\wedge (\odot
^q\frkg_{-1}),V_{-1})\longrightarrow\Hom((\wedge  ^p\frkg_0)\wedge
(\odot ^q\frkg_{-1}),V_{0})$  defined by$$
  \hat{\partial}(f)=(-1)^{p+2q}\partial\circ f;
$$
The operator $d_\mu$ can be written as
$d_\mu=(d_\mu^{(1,0)},d_\mu^{(0,1)})$, where

$\bullet$ \, $d_\mu^{(1,0)}:\Hom((\wedge  ^p\frkg_0)\wedge (\odot
^q\frkg_{-1}),V_s)\longrightarrow \Hom((\wedge ^{p+1}\frkg_0)\wedge
(\odot  ^q\frkg_{-1}),V_s)$ defined by
\begin{eqnarray*}
&& d_\mu^{(1,0)}(f)(x_1,\cdots,x_{p+1},a_1,\cdots,a_{q})\\
 &=&\sum_{i=1}^{p+1}(-1)^{i+1}\mu_0(x_i)f(x_1,\cdots,\widehat{x_i},\cdots,x_{p+1},
  a_1,\cdots,a_{q})\\
  &&+\sum_{i<j}(-1)^{i+j}f([x_i,x_j]_\g,x_1,\cdots,\widehat{x_i},\cdots,\widehat{x_j}\cdots,x_{p+1},
  a_1,\cdots,a_{q})\\
  &&+\sum_{i,j}(-1)^{i}f(x_1,\cdots,\widehat{x_i},\cdots,x_{p+1},
  a_1,\cdots,[x_i,a_j]_\g,\cdots,a_{q});
  \end{eqnarray*}

$\bullet$\, $d_\mu^{(0,1)}:\Hom((\wedge  ^p\frkg_0)\wedge (\odot
^q\frkg_{-1}),V_0)\longrightarrow \Hom((\wedge  ^p\frkg_0)\wedge
(\odot  ^{q+1}\frkg_{-1}),V_{-1})$ defined by
\begin{eqnarray*}
 d_\mu^{(0,1)}(f)(x_1,\cdots,x_{p},a_1,\cdots,a_{q+1})
 &=&\sum_{i=1}^{q+1}(-1)^{p}\mu_1(a_i)f(x_1,\cdots,x_{p},
  a_1,\cdots,\widehat{a_i},\cdots,a_{q+1});
\end{eqnarray*}

$\bullet$ \, $d_{\mu_2}:\Hom((\wedge^{p}\frkg_0)\wedge
\odot^q\frkg_{-1},V_0)\longrightarrow
\Hom((\wedge^{p+2}\frkg_0)\wedge \odot^q\frkg_{-1},V_{-1})$ defined
by
$$
  d_{\mu_2}f(x_1,\cdots,x_{p+2},a_1,\cdots,a_{q})=
  \sum_{\sigma}(-1)^{p+2q}(-1)^\sigma\mu_2(x_{\sigma(1)},x_{\sigma(2)})f(x_{\sigma(3)},
  \cdots,x_{\sigma(p+2)},a_1,\cdots,a_{q});
$$

$\bullet$ \, $d_{l_3}:\Hom((\wedge^p\frkg_0)\wedge \odot^q
\frkg_{-1},V_s)\longrightarrow \Hom((\wedge^{p+3}\frkg_0)\wedge
\odot^{q-1} \frkg_{-1},V_s)$ defined by
\begin{eqnarray*}
&&d_{l_3^\g}f(x_1,\cdots,x_{p+3},a_1,\cdots,a_{q-1})\\
&&=\sum_\sigma-(-1)^\sigma f(x_{\sigma(4)},\cdots,x_{\sigma(p+3)},a_1,\cdots,a_{q-1},l_3^\g(x_{\sigma(1)},x_{\sigma(2)},x_{\sigma(3)})).
\end{eqnarray*}
Thus, one can  write the cochain  complex in components as follows:

{\footnotesize
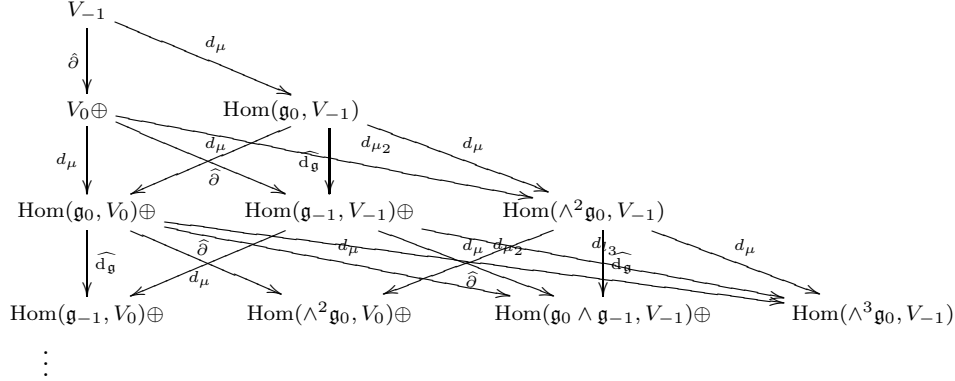
\xymatrix{
  V_{-1} \ar[d]_{\hat{\partial}} \ar[dr]^{d_{\mu}} & & & & \\
  V_{0}\oplus \ar[d]_{d_{\mu}}\ar[dr]^ {d_{\mu}} \ar[drr]^{\qquad d_{\mu_2}} &\hspace{-1cm} \Hom(\frkg_{0},V_{-1})
          \ar[dl]^{\widehat{\partial}}\ar[d]_{\widehat{\dM_\g}}\ar[dr]^{d_{\mu}}&&& \\
  \Hom(\frkg_{0},V_{0})\oplus\ar[d]^{\widehat{\dM_\g}}\ar[dr]_{d_{\mu}}\ar[drr]^{d_{\mu}}\ar[drrr]^{\qquad d_{\mu_2}}  &\Hom(\frkg_{-1},V_{-1})\oplus
            \ar[dl]_{\widehat{\partial}} \ar[dr]^{d_{\mu}} \ar[drr]^{d_{l_3}}     & \hspace{-.5cm}\Hom(\wedge^2\frkg_0,V_{-1})
            \ar[dl]^{\widehat{\partial}} \ar[d]^{\widehat{\dM_\g}} \ar[dr]^{d_{\mu}}  &&\\
 \Hom(\frkg_{-1},V_{0}) \oplus
      & \Hom(\wedge^2\frkg_0,V_{0}) \oplus
       &\Hom(\frkg_0\wedge \frkg_{-1},V_{-1}) \oplus
      &\Hom(\wedge^3\frkg_0,V_{-1})
       &  \\
        }}

        $\vdots$

For a 2-cochain $(\psi,\omega,\nu,\theta)$, where $\psi\in
\Hom(\frkg_{-1},V_{0}),\ \ \omega\in\Hom(\wedge^2\frkg_{0},V_{0}),
 \ \ \nu\in\Hom(\frkg_{0}\wedge\frkg_{-1},V_{-1}), \ \ \theta\in \Hom(\wedge^3\frkg_{0},V_{-1})$, it is a 2-cocycle if and only if
 the following equations hold:
 \begin{eqnarray*}
   (d_\mu\psi+\hat{\dM_\g}\omega+\hat{\partial}\nu)(x,a)&=&0,\\
   (d_\mu\psi+\hat{\dM_\g}\nu)(a,b)&=&0,\\
   (d_{l_3}\psi+d_\mu\omega+\widehat{\partial}\theta)(x,y,z)&=&0,\\
   (d_{\mu_2}\psi+d_\mu\omega+d_\mu\nu+\hat{\dM_\g}\theta)(x,y,a)&=&0,\\
   (d_{\mu_2}\omega+d_{l_3}\nu+d_\mu\theta)(x,y,z,t)&=&0.
 \end{eqnarray*}
 More precisely,
\begin{eqnarray}
\label{eq:2-cocycle01}\mu_0(x)\psi(a)-\psi([x,a]_\g)+\omega(x,\dM_\g a)-\partial\circ\nu(x,a)&=&0,\\
\label{eq:2-cocycle02}\mu_1(a)\psi(b)+\mu_1(b)\psi(a)-\nu(\dM_\g b,a)-\nu(\dM_\g a,b)&=&0,\\
\label{eq:2-cocycle1} -\psi(l_3^\g(x,y,z))+\mu_0(x)\omega(y,z)+c.p.-\omega([x,y]_\g,z)+c.p.-\partial\circ\theta(x,y,z)&=&0,\\
\nonumber\mu_1(a)\omega(x, y)+\mu_0(x)\nu(y,a)-\mu_0(y)\nu(a,x)-\nu([x,y]_\g,a)+\nu([x,a]_\g,y)-\nu([y,a]_\g,x)\\
\label{eq:2-cocycle2}-\theta(x,y,\dM_\g a)+\mu_2(x,y)(\psi (a))&=&0,\\
\label{eq:2-cocycle3}(\mu_2(z,t)(\omega(x,y))-\nu(t,l_3^\g(x,y,z))+\mu_0(x)\theta(y,z,t)-\theta([x,y]_\g,z,t))+c.p.&=&0.
\end{eqnarray}

For a Lie 2-algebra $(\frkg;\dM_\g,[\cdot,\cdot]_\g,l_3^\g)$, there
is a natural {\bf adjoint representation} on itself. The
corresponding $(\mu_0,\mu_1,\mu_2)$, which we denote by
$\ad=(\ad^0,\ad^1,\ad^2)$ now,  is given by
$$
\ad^0_x(y+b)=[x,y]_\g+[x,b]_\g,\quad \ad^1_ax=[a,x]_\g,\quad \ad^2_{x,y}z=-l_3^\g(x,y,z).
$$
We can view the Lie 2-algebra structure $(\dM_\g,[\cdot,\cdot]_\g,l_3^\g)$ as a 2-cochain satisfying some conditions.

\begin{ex}\label{ex:1}{\rm
 Let $(\s,[\cdot,\cdot]_\s,\langle\cdot,\cdot\rangle_\s)$ be a quadratic Lie algebra. It gives a Lie 2-algebra structure on $\s\oplus\R$, where $\s$ is of degree 0, and $\R$ is of degree $-1$, and  $\dM,[\cdot,\cdot],l_3$ are given by
$$
\dM=0,\quad [x+a,y+b]=[x,y]_\s,\quad l_3(x,y,z)=\langle [x,y]_\s,z\rangle_\s.
$$
We assume that $l_3\neq0$, and denote the corresponding Lie 2-algebra by  $\Lie(\s)$. If $\s$ is semisimple and  $\langle\cdot,\cdot\rangle_\s$ is the Killing form on $\s$, we obtain the string Lie 2-algebra $\str(\s)$.

Since $\Hom(\wedge^k\s,\R)=\wedge^k\s^*$, the generalized
Chevalley-Eilenberg complex of $\Lie(\s)$ associated to the adjoint representation is given by
\begin{equation} \label{eq:complex}
\begin{split}
 & \R\stackrel{D}{\longrightarrow}
  \s\oplus \s^*\stackrel{D}{\longrightarrow}
  \Hom(\s,\s)\oplus\R\oplus\wedge^2\s^*\stackrel{D}{\longrightarrow}\\
  & \Hom(\R,\s)\oplus \Hom(\wedge^2\s,\s)\oplus \Hom(\s\wedge \R,\R) \oplus \wedge^3\s^*\stackrel{D}{\longrightarrow}\cdots,
\end{split}
\end{equation}
It is easy to show that the nonzero components of $D$ is only $d^{(1,0)}_\ad,~d_{\ad^2}$ and $d_{l_3}$. Denote by $\HH^k(\s;\ad)$ and $\HH^k(\s)$ the $k$-th cohomology group of $\s$ with coefficients in the adjoint representation and the trivial representation respectively.

\begin{itemize}
  \item[$\bullet$] For all $(-1)$-cochain $a \in \R$, we have $D(a)=0$. So we have $\HH^{-1}(\Lie(\s);\ad)=\R$;

   \item[$\bullet$] For all $(x,\xi)\in\s\oplus\s^*$, if $D(x,\xi)=0$, consider the components in $\Hom(\s,\s)$, we have $[x,y]_\s=0$ for all $y\in\s$, which implies that $x\in \Cen(\s)$, where $\Cen(\s)$ is the center of $\s$; the component in $\R$ is $0$ naturally; consider the component in $\wedge^2\s^*$, we have $-l_3(y,z,x)-\xi([y,z]_\s)\rangle=0$. Since $x\in\Cen(\s)$, we obtain that $\xi([y,z]_\s)=0$. Thus, we have  $$\HH^{0}(\Lie(\s);\ad)=\HH^0(\s;\ad)\oplus\HH^1(\s).$$
\end{itemize}
}

\end{ex}

\section{Deformations of Lie 2-algebras}

\subsection{1-parameter infinitesimal deformations of Lie 2-algebras}

Let $(\frkg, \dM_\g, [\cdot,\cdot]_\g, l_3^\g)$ be a Lie 2-algebra, and $\omega_1:\g_{-1}\longrightarrow \g_0,~\omega_2^0:\wedge^2\g_{0}\longrightarrow \g_0,~\omega_2^1:\g_0\wedge\g_{-1}\longrightarrow \g_{-1},~\omega_3:\wedge^3\g_0\longrightarrow\g_{-1}$ be linear maps. Consider a $\lambda$-parametrized family of linear operations:
\begin{eqnarray*}
 \dM^\lambda a&\triangleq&\dM_\g a+\lambda\omega_1(a),\\
 %l_2^\lambda (x, y)=
 {[x,y]}\dlam&\triangleq& [x,y]_\g+ \lambda\omega^0_2(x, y),\\
 %l_2^\lambda (x, a)=
 {[x,a]}\dlam&\triangleq& [x,a]_\g+ \lambda\omega^1_2(x, a),\\
 l_3^\lambda (x, y, z)&\triangleq& l_3^\g(x, y, z)+ \lambda\omega_3(x, y, z).
 \end{eqnarray*}
%\comment{Since we use $[\cdot,\cdot]$ at the beginning, here we need to unify.}

If all $(\dM^\lambda, [\cdot,\cdot]\dlam, l^\lambda_3)$ endow the graded vector space $\g_0\oplus\g_{-1}$ with  Lie 2-algebra structures, we say that
$(\omega_0,\omega^0_2,\omega^1_2,\omega_3)$ generates a 1-parameter infinitesimal deformation of the Lie 2-algebra $\frkg$.

\begin{thm}\label{thm:deformation}
$(\omega_0,\omega^0_2,\omega^1_2,\omega_3)$ generates a $1$-parameter infinitesimal deformation of the Lie $2$-algebra $\frkg$ is equivalent to the following conditions:
\begin{itemize}
  \item[\rm(i)] $(\omega_1,\omega^0_2,\omega^1_2,\omega_3)$ is a $2$-cocycle of $\frkg$ with the coefficients in the adjoint representation;

  \item[\rm(ii)] $(\omega_1,\omega^0_2,\omega^1_2,\omega_3)$ itself defines a Lie $2$-algebra structure on $\g_0\oplus\g_{-1}$.
\end{itemize}
\end{thm}

\pf
If $(\frkg, \dM^\lambda, [\cdot,\cdot]\dlam, l^\lambda_3)$ is a Lie 2-algebra, by (i) in Definition \ref{defi:Lie 2}, we have
\begin{eqnarray*}
%&&d\ulam l\ulam_2(x,a)-l\ulam_2(x, d\ulam a)\\
%&=&(d+\lambda\omega_1)(l_2+\lambda\omega^1_2)(x,a)-(l_2+\lambda\omega^1_2)(x, (d+\lambda\omega_1)a)\\
%&=&dl_2(x,a)+\lambda(\omega_1l_2+d\omega^1_2)(x,a)+\lambda^2\omega_1\omega^0_2(x,a)\\
%&&-l_2(x,d(a))-\lambda(\omega^1_2(x, da)+l_2(x,\omega_1a))-\lambda^2\omega^1_2(x,\omega_1a)\\
&&\dM\ulam [x,a]\dlam-[x, \dM\ulam a]\dlam\\
&=&(\dM_\g+\lambda\omega_1)([x,a]_\g+ \lambda\omega^1_2(x, a))-[x, \dM_\g a+\lambda\omega_1a]_\g-\lambda\omega^0_2(x, \dM_\g a+\lambda\omega_1a)\\
&=&\dM_\g[x,a]_\g+\lambda(\omega_1[x,a]_\g+\dM_\g\omega^1_2(x,a))+\lambda^2\omega_1\omega^0_2(x,a)\\
&&-[x,\dM_\g a]_\g-\lambda(\omega^0_2(x, \dM _\g a)+[x,\omega_1a]_\g)-\lambda^2\omega^1_2(x,\omega_1a)\\
&=&0,
\end{eqnarray*}
%\comment{You see you made many mistake in $\omega^0_2$ and $\omega^1_2$, here and through the paper}
which implies that
\begin{eqnarray}
\label{eq:2-cocycle01'}\omega_1[x,a]_\g+\dM_\g \omega_2^1(x,a)-\omega^0_2(x, \dM_\g a)-[x,\omega_1a]_\g&=&0,\\
\label{eq:2-cocycle01''}\omega_1\omega^1_2(x,a)-\omega^0_2(x,\omega_1a)&=&0.
\end{eqnarray}

Similarly, by (ii) in Definition \ref{defi:Lie 2}, we have
%\begin{eqnarray*}
%&&l\ulam_2(d\ulam a,b)-l\ulam_2(x, d\ulam b)\\
%&=&(l_2+\lambda\omega_2)((l_1+\lambda\omega_1)a,b)-(l_2+\lambda\omega_2)(a, (l_1+\lambda\omega_1)b)\\
%&=&l_2(l_1a,b)+\lambda(l_2(\omega_1a,b)+\omega^1_2(l_1a,b))+\lambda^2\omega^1_2(\omega_1a,b)\\
%&&-l_2(a,l_1(b))-\lambda(\omega^1_2(a, db)+[a,\omega_1b])-\lambda^2\omega^1_2(a,\omega_1b)\\
%&=&0,
%\end{eqnarray*}
\begin{eqnarray*}
&&[\dM\ulam a,b]\dlam-[a, \dM\ulam b]\dlam\\
&=&[\dM_\g a+\lambda\omega_1a,b]_\g+\lambda\omega_2^1(\dM_\g a+\lambda\omega_1a,b)-[a,\dM_\g b+\lambda\omega_1b]_\g-\lambda\omega_2^1(a, \dM_\g b+\lambda\omega_1b)\\
&=&[\dM_\g a,b]_\g+\lambda([\omega_1a,b]_\g+\omega^1_2(\dM_\g a,b))+\lambda^2\omega^1_2(\omega_1a,b)\\
&&-[a,\dM_\g b]_\g-\lambda(\omega^1_2(a, \dM_\g b)+[a,\omega_1b]_\g)-\lambda^2\omega^1_2(a,\omega_1b)\\
&=&0,
\end{eqnarray*}
which implies that
\begin{eqnarray}
\label{eq:2-cocycle02'}[\omega_1a,b]_\g-[a,\omega_1b]_\g+\omega^1_2(\dM_\g a,b)-\omega^1_2(a,\dM_\g b)&=&0,\\
\label{eq:2-cocycle02''}\omega^1_2(\omega_1a,b)-\omega^1_2(a,\omega_1b)&=&0.
\end{eqnarray}

By  (iii) in Definition \ref{defi:Lie 2}, we have
\begin{eqnarray*}
&&[[x,y]\dlam,z]\dlam+c.p.+\dM\ulam l\ulam_3(x,y,z)\\
&=&[[x, y]_\g+\lambda\omega^0_2(x, y), z]\dlam+c.p.+(\dM_\g +\lambda\omega_1)(l_3^\g+\lambda\omega_3)(x,y,z)\\
&=&[[x, y]_\g,z]\dlam+c.p.+\lambda[\omega^0_2(x, y),z]\dlam+c.p.+(\dM_\g l_3^\g+\lambda(\omega_1l_3^\g+\dM_\g \omega_3)+\lambda^2\omega_1\omega_3)(x,y,z)\\
&=&[[x, y]_\g,z]_\g+c.p.+\lambda(\omega^0_2([x,y]_\g,z)+[\omega^0_2(x, y),z]_\g)+c.p.+\lambda^2\omega^0_2(\omega^0_2(x,y),z)+c.p.\\
&&+\dM_\g l_3^\g(x,y,z)+\lambda(\omega_1l_3^\g+\dM_\g \omega_3)(x,y,z)+\lambda^2\omega_1\omega_3(x,y,z),\\
&=&0,
\end{eqnarray*}
which implies that
\begin{eqnarray}
\label{eq:2-cocycle1'}\omega^0_2([x,y]_\g,z)+c.p.+[\omega^0_2(x, y),z]_\g+c.p.+(\omega_1l_3^\g+\dM_\g \omega_3)(x,y,z)&=&0,\\
\label{eq:2-cocycle1''}\omega^0_2(\omega^0_2(x,y),z)+c.p.+\omega_1\omega_3(x,y,z)&=&0.
\end{eqnarray}

By  (iv) in Definition \ref{defi:Lie 2}, we have
\begin{eqnarray*}
&&[[x,y]\dlam,a]\dlam+c.p.+ l\ulam_3(x,y,\dM \ulam a)\\
&=&[[x, y]_\g+\lambda\omega^0_2(x, y), a]\dlam+c.p.+(l_3^\g+\lambda\omega_3)(x,y,(\dM_\g +\lambda\omega_1)a)\\
&=&[[x, y]_\g,a]\dlam+c.p.+\lambda[\omega^0_2(x, y),a]\dlam+c.p.+l_3^\g(x,y,(\dM_\g +\lambda\omega_1)a)+\lambda\omega_3(x,y,(\dM_\g +\lambda\omega_1)a)\\
&=&[[x, y]_\g,a]_\g+c.p.+\lambda(\omega^1_2([x,y]_\g,a)+[\omega^0_2(x, y),a]_\g)+c.p.+\lambda^2\omega^1_2(\omega^0_2(x,y),a)+c.p.\\
&&+l_3^\g(x,y,\dM_\g a)+\lambda(l_3^\g(x,y,\omega_1a)+\omega_3(x,y,\dM_\g a))+\lambda^2\omega_3(x,y,\omega_1a),\\
&=&0,
\end{eqnarray*}
which implies that
\begin{eqnarray}
\label{eq:2-cocycle2'}&&\nonumber\omega^1_2([x,y]_\g,a)+\omega^1_2([y,a]_\g,x)+\omega^1_2([a,x]_\g,y)
+[\omega^0_2(x, y),a]_\g+[\omega^1_2(y, a),x]_\g+[\omega^1_2(a, x),y]_\g\\
&&+l_3^\g(x,y,\omega_1a)+\omega_3(x,y,\dM_\g a)=0,\\
%\label{eq:2-cocycle2''}\omega^1_2(\omega^0_2(x,y),a)+c.p.+\omega_3(x,y,\omega_1a)&=&0.
\label{eq:2-cocycle2''}&&\omega^1_2(\omega^0_2(x,y),a)+\omega^1_2(\omega^1_2(y,a),x)+\omega^1_2(\omega^1_2(a,x),y)+\omega_3(x,y,\omega_1a)=0.
\end{eqnarray}

For the equality
%$$l\ulam_3(l\ulam_2(x,y),z,t)+c.p.-l\ulam_2(l\ulam_3(x,y,z),t)-c.p.=0,$$
$$l\ulam_3([x,y]\dlam,z,t)+c.p.-[l\ulam_3(x,y,z),t]\dlam-c.p.=0,$$
we have
\begin{eqnarray*}
&&l\ulam_3([x,y]\dlam,z,t)+c.p.-[l\ulam_3(x,y,z),t]\dlam-c.p.\\
&=&(l_3^\g+\lambda\omega_3)([x,y]_\g,z,t)+c.p.+\lambda(l_3^\g+\lambda\omega_3)(\omega^0_2(x,y),z,t)+c.p.\\
&&-[l_3^\g(x,y,z),t]\dlam-c.p.-\lambda[\omega_3(x,y,z),t]\dlam-c.p.\\
&=&l_3^\g([x,y]_\g,z,t)+c.p.+\lambda\omega_3([x,y]_\g,z,t)+c.p.\\
&&+\lambda l_3^\g(\omega^0_2(x,y),z,t)+c.p.+\lambda^2\omega_3(\omega^0_2(x,y),z,t)+c.p.\\
&&-[l_3^\g(x,y,z),t]_\g-c.p.-\lambda\omega^1_2(l_3^\g(x,y,z),t)-c.p.\\
&&-\lambda [\omega_3(x,y,z),t]_\g-c.p.-\lambda^2\omega^1_2(\omega_3(x,y,z),t)-c.p.\\
&=&0,
\end{eqnarray*}
which implies that
\begin{eqnarray}
\label{eq:2-cocycle3'} (\omega_3([x,y]_\g,z,t)+l_3^\g(\omega^0_2(x,y),z,t))+c.p.&=&(\omega^1_2(l_3^\g(x,y,z),t)+ [\omega_3(x,y,z),t]_\g)+c.p.,\\
\label{eq:2-cocycle3''}\omega_3(\omega^0_2(x,y),z,t)+c.p.&=&\omega^1_2(\omega_3(x,y,z),t)+c.p..
\end{eqnarray}
By \eqref{eq:2-cocycle01'}, \eqref{eq:2-cocycle02'}, \eqref{eq:2-cocycle1'}, \eqref{eq:2-cocycle2'} and \eqref{eq:2-cocycle3'}, we deduce that $(\omega_1,\omega_2^0,\omega_2^1,\omega_3)$ is a 2-cocycle of $\g$ with the coefficients in the adjoint representation.

Furthermore, by \eqref{eq:2-cocycle01''}, \eqref{eq:2-cocycle02''}, \eqref{eq:2-cocycle1''}, \eqref{eq:2-cocycle2''} and \eqref{eq:2-cocycle3''},
 $(\g;\omega_1, \omega^0_2, \omega^1_2, \omega_3)$ is a Lie 2-algebra.
\qed

\subsection{Nijenhuis operators}

In this subsection, we introduce the notion of Nijenhuis operators,
which could give trivial deformations.

Let $(\frkg, \dM_\g, [\cdot,\cdot]_\g, l_3^\g)$ be a Lie 2-algebra, and $N=(N_0,N_1)$ be a chain map, i.e. $N_0:\frkg_0\to \frkg_0$ and $N_1: \frkg_{-1}\to \frkg_{-1}$
 are  linear maps satisfying $\dM_\g\circ N_1=N_0\circ \dM_\g$. Define an exact 2-cochain
 $$(\dM_N,[\cdot,\cdot]_N,l_3^N)=D(N_0,N_1).$$
 by differential $D$ discussed in Section 2, i.e.,
  \begin{eqnarray*}
   \dM_N&=&\dM_\g\circ N_1-N_0\circ \dM_\g=0,\\
~[x, y]_N&=&[N_0x, y]_\g + [x, N_0y]_\g - N_0[x, y]_\g,\\
~[x, a]_N&=&[N_0x, a]_\g + [x, N_1a]_\g - N_1[x, a]_\g,\\
l_3^N(x,y,z)&=&l_3^\g(N_0x,y,z)+c.p.-N_1(l_3^\g(x,y,z)).
 \end{eqnarray*}

 \begin{defi}\label{defi:Nijenhuis}
  A chain map $N=(N_0,N_1)$ is called a Nijenhuis operator if for all $x,y,z\in\g_0$ and $a\in\g_{-1}$, the following conditions are satisfied:
   \begin{itemize}
     \item[(i)] $\dM_\g\circ N_1=N_0\circ \dM_\g =0;$
     \item[(ii)] $N_0[x,y]_N=[N_0x,N_0y]_\g;$
     \item[(iii)] $N_1[x,a]_N=[N_0x,N_1a]_\g;$
     \item[(iv)] $ N_1l_3^N(x,y,z)=0;$
     \item[(v)] $l_3^\g(N_0x,N_0y,N_0z)=0;$
     \item[(vi)] $l_3^\g(N_0x,N_0y,z)+c.p.=0.$
   \end{itemize}
 \end{defi}

 \begin{pro}\label{pro:Nijenhuis}
   Let $N=(N_0,N_1)$ be a Nijenhuis operator, then $\lambda N=(\lambda N_0,\lambda  N_1)$ is also a Nijenhuis operator for all
    $\lambda\in\R$. Furthermore,  $(\dM_{\lambda N}=0,[\cdot,\cdot]_{\lambda N},l_3^{\lambda N})$ defines
     a skeletal Lie $2$-algebra structure on $\g$, and
 $$\lambda N: (\g;\dM_{\lambda N}=0,[\cdot,\cdot]_{\lambda N},l_3^{\lambda N}) \longrightarrow
     (\g;\dM_\g,[\cdot,\cdot]_\g,l_3^\g)$$
is a homomorphism of Lie $2$-algebras.
 \end{pro}
\pf It is obvious that conditions (i)-(vi) holds for $\lambda N$,
which implies that $\lambda N$ is a Nijenhuis operator, for all
$\lambda.$ To prove that $(\dM_{\lambda N}=0,[\cdot,\cdot]_{\lambda
N},l_3^{\lambda N})$ defines a skeletal Lie $2$-algebra structure on
$\g$ for all $\lambda$, it is sufficient to show the case that
$\lambda=1$. By the Nijenhuis condition (ii), we have
\begin{eqnarray*}
 [ [x,y]_N,z]_N+c.p.=\dM_\g \big(l_3^\g(N_0x,N_0y,z)+c.p.\big)  +N_0\circ \dM_\g \big(l_3^\g(N_0x,y,z)+c.p.\big) +N_0^2\circ \dM_\g l_3^\g(x,y,z).
\end{eqnarray*}
By the Nijenhuis conditions (i) and (vi), we deduce that $[ [x,y]_N,z]_N+c.p.=0.$ Similarly, we can prove that $[ [x,y]_N,a]_N+c.p.=0.$

Furthermore, by the Nijenhuis conditions (iii), (iv) and (vi), we have
\begin{eqnarray*}
  J_{l_3^N}(x,y,z,t)=J_{l_3^\g}(N_0x,N_0y,z,t)+c.p.+N_1J_{l_3^\g}(N_0x,y,z,t)+c.p+N_1^2J_{l_3^\g}(x,y,z,t)=0.
\end{eqnarray*}
Therefore, $(\g,\dM_N=0,[\cdot,\cdot]_N,l_3^N)$ is a skeletal Lie 2-algebra.

To see that $\lambda N$ is a homomorphism, first the Nijenhuis condition (i) guarantees that $N$ commutes with the differential. Since the corresponding $F_2$ in Definition \ref{defi:Lie-2hom} is zero here, it is obvious that conditions (ii) and (iii) in Definition \ref{defi:Lie-2hom} hold. At last, the condition (iv) in Definition \ref{defi:Lie-2hom} reduce to $\lambda N_1l_3^{\lambda N}(x,y,z)=l_3^\g(\lambda N_0x,\lambda N_0y,\lambda N_0z)$, which holds by the Nijenhuis conditions (iv) and (v). \qed\vspace{3mm}

In the following, we show that for all polynomial $P$, $P(N)$ is
also a Nijenhuis operator. To do that, we need some preparation.

\begin{lem}
Let $N=(N_0,N_1)$ be a Nijenhuis operator, then for all $j,k>0$, we
have
\begin{eqnarray}
 \label{eq:Pc1} N_1^jl_3^{N^k}(x,y,z)&=&0,\\
 \nonumber l_3^\g(N^k_0x,N^j_0y,z)+l_3^\g(N^k_0x,y,N^j_0z)+l_3^\g(x,N^k_0y,N^j_0z)\\
\label{eq:Pc2}+l_3^\g(N^j_0x,N^k_0y,z)+l_3^\g(N^j_0x,y,N^k_0z)+l_3^\g(x,N^j_0y,N^k_0z)&=&0.
\end{eqnarray}
\end{lem}
\pf
We prove $N_1l^{N^k}_3(x,y,z)=0$ by induction. Assume that $N_1l^{N^{k-1}}_3(x,y,z)=0$, i.e.
$$N_1l_3^\g(N^{k-1}_0x,y,z)+N_1l_3^\g(x,N^{k-1}_0y,z)+N_1l_3^\g(x,y,N^{k-1}_0z)=N_1N^{k-1}_1l_3^\g(x,y,z).$$
Substitute $x,y,z$ by $N^{k-1}_0x$, $N^{k-1}_0y$, $N^{k-1}_0z$ respectively in the equation $N_1l^N_3(x,y,z)=0$, we obtain
\begin{eqnarray*}
N_1l_3^\g(N^k_0x,y,z)+N_1l_3^\g(N^{k-1}_0x,N_0y,z)+N_1l_3^\g(N^{k-1}_0x,y,N_0z)-N^2_1l_3^\g(N^{k-1}_0x,y,z)=0,\\
N_1l_3^\g(N_0x,N^{k-1}_0y,z)+N_1l_3^\g(x,N^k_0y,z)+N_1l_3^\g(x,N^{k-1}_0y,N_0z)-N^2_1l_3^\g(x,N^{k-1}_0y,z)=0,\\
N_1l_3^\g(N_0x,y,N^{k-1}_0z)+N_1l_3^\g(x,N_0y,N^{k-1}_0z)+N_1l_3^\g(x,y,N^k_0z)-N^2_1l_3^\g(x,y,N^{k-1}_0z)=0.
\end{eqnarray*}
The sum of the left hand side of the above three equations are
\begin{eqnarray*}
&&N_1l_3^\g(N^k_0x,y,z)+N_1l_3^\g(x,N^k_0y,z)+N_1l_3^\g(x,y,N^k_0z)\\
&&-N^2_1(l_3^\g(N^{k-1}_0x,y,z)+l_3^\g(x,N^{k-1}_0y,z)+l_3^\g(x,y,N^{k-1}_0z))\\
&=&N_1l_3^\g(N^k_0x,y,z)+N_1l_3^\g(x,N^k_0y,z)+N_1l_3^\g(x,y,N^k_0z)-N_1N^{k}_1l_3^\g(x,y,z)\\
&=&N_1l^{N^k}_3 (x,y,z).
\end{eqnarray*}
Thus, we have $N_1l^{N^k}_3(x,y,z)=0$, and $N^{j}_1l^{N^k}_3(x,y,z)=N^{j-1}_1(N_1l^{N^k}_3(x,y,z))=0$. This finishes the proof of \eqref{eq:Pc1}.

For all $k>1$, we have
$$l_3^\g(N^k_0x,N^k_0y,z)=l_3^\g(N^k_0x,N^k_0y,z)+l_3^\g(N^k_0x,N^{k-1}_0y,N_0z)+l_3^\g(N^{k-1}_0x,N^k_0y,N_0z)=0.$$
Therefore, for all $k>0$, we have
$$l_3^\g(N^k_0x,N^k_0y,z)+l_3^\g(N^k_0x,y,N^k_0z)+l_3^\g(x,N^k_0y,N^k_0z)=0.$$
Without loss of generality, we assume that $j>k$, and  substitute $x,y,z$ by $N^{j-k}_0x$, $N^{j-k}_0y$, $N^{j-k}_0z$ in the above equation respectively, we get
\begin{eqnarray*}
&&l_3^\g(N^j_0x,N^k_0y,z)+l_3^\g(N^j_0x,y,N^k_0z)+l_3^\g(N^{j-k}_0x,N^k_0y,N^k_0z)=0,\\
&&l_3^\g(N^k_0x,N^j_0y,z)+l_3^\g(N^k_0x,N^{j-k}_0y,N^k_0z)+l_3^\g(x,N^j_0y,N^k_0z)=0,\\
&&l_3^\g(N^k_0x,N^k_0y,N^{j-k}_0z)+l_3^\g(N^k_0x,y,N^j_0z)+l_3^\g(x,N^k_0y,N^j_0z)=0.
\end{eqnarray*}
Then we obtain \eqref{eq:Pc2} by considering the sum of the left hand side of the above three equations, and the fact that
$$l_3^\g(N^{j-k}_0x,N^k_0y,N^k_0z)=l_3^\g(N^k_0x,N^{j-k}_0y,N^k_0z)=l_3^\g(N^k_0x,N^k_0y,N^{j-k}_0z)=0.$$
The proof is completed.\qed

\begin{thm}
  Let $N = (N_0,N_1)$ be a Nijenhuis operator, then for all polynomial
$P(X)=\sum_{i=1}^n c_i X^i$, the operator $P(N)=(P(N_0),P(N_1))$ is also a Nijenhuis operator.
\end{thm}

\pf
The Nijenhuis conditions (i) and (v) hold obviously.
We can also prove that the Nijenhuis conditions (ii) and (iii) hold similar as the proof of \cite[Proposition 3.3]{D}, and we do not repeat it here.  By \eqref{eq:Pc2}, we have
\begin{eqnarray*}
P(N_1)l^{P(N)}_3(x,y,z)=\sum_{j,k=1}^n c_jc_kN^j_1l^{N^k}_3(x,y,z)=0,
\end{eqnarray*}
which implies that the Nijenhuis condition (iv) hold.
By \eqref{eq:Pc1},  we have
\begin{eqnarray*}
l_3^\g(P(N_0) x, P(N_0) y, z)+c.p.=\sum_{j,k=1}^n c_jc_kl_3^\g(N^j_0 x, N^k_0 y, z)+c.p.=0,
\end{eqnarray*}
which implies that the Nijenhuis condition (vi) holds. This
completes the proof.\qed\vspace{3mm}

\emptycomment{\begin{pro}
  Let $N=(N_0,N_1)$ be a Nijenhuis operator, then $N^2=(N_0^2,N_1^2)$ is also a Nijenhuis operator.
\end{pro}
\pf The Nijenhuis conditions (i)-(iii) and (v) hold obviously. To see the Nijenhuis condition (iv) holds, we need to prove that
$$
N_1^2(l_3^\g(N_0^2x,y,z)+c.p.)-N_1^4l_3^\g(x,y,z)=0.
$$
Since $(N_1,N_1)$ is a Nijenhuis operator, we have
\begin{eqnarray*}
 N_1^4l_3^\g(x,y,z)&=&N_1^3(l_3^\g(N_0x,y,z)+l_3^\g(x,N_0y,z)+l_3^\g(x,y,N_0z))\\
 &=& N_1^2(l_3^\g(N_0^2x,y,z)+l_3^\g(N_0x,N_0y,z)+l_3^\g(N_0x,y,N_0z))\\
 &&+N_1^2(l_3^\g(N_0x,N_0y,z)+l_3^\g(x,N_0^2y,z)+l_3^\g(x,N_0y,N_0z))\\
 &&+ N_1^2(l_3^\g(N_0x,y,N_0z)+l_3^\g(x,N_0y,N_0z)+l_3^\g(x,y,N_0^2z))\\
 &=&N_1^2(l_3^\g(N_0^2x,y,z)+c.p.)+2N_1^2(l_3^\g(N_0x,N_0y,z)+c.p.)\\
 &=&N_1^2(l_3^\g(N_0^2x,y,z)+c.p.).
\end{eqnarray*}
The last equality hold by   the ~Nijenhuis~
  condition ~(vi). Thus, the Nijenhuis condition (iv) hold for $N^2$.

  To see the Nijenhuis condition (vi) hold for $N^2$, we need to prove that
  $$
  l_3^\g(N_0^2x,N_0^2y,z)+c.p.=0,
  $$
which follows from
$$
0=l_3^\g(N_0^2x,N_0^2y,z)+l_3^\g(N_0^2x,N_0y,N_0z)+l_3^\g(N_0x,N_0^2y,N_0z)=l_3^\g(N_0^2x,N_0^2y,z).
$$
This completes the proof. \qed}

 Let $(\s,[\cdot,\cdot]_\s,\langle\cdot,\cdot\rangle_\s)$ be the Lie 2-algebra $\Lie(\s)$ given in Example \ref{ex:1}.
 As a special case, we consider  operator
    $N_0:\s\longrightarrow\s,~N_1= 0:\R\longrightarrow \R$. By Definition \ref{defi:Nijenhuis}, it is easy to check that $N=(N_0, 0)$ is a Nijenhuis operator if and only if
the following equalities hold:
\begin{eqnarray}
\label{eq:SNijenhuis01}[N_0x, N_0y]_\s - N_0[N_0x, y]_\s - N_0[x, N_0y]_\s + N_0^2[x,y]_\s&=&0,\\
\label{eq:SNijenhuis02}\langle[N_0x,N_0y]_\s,N_0z\rangle_\s&=&0,\\
\label{eq:SNijenhuis03}\langle[N_0x,N_0y]_\s,z\rangle_\s+\langle[N_0x,y]_\s,N_0z\rangle_\s+\langle[x,N_0y]_\s,N_0z\rangle_\s&=&0.
\end{eqnarray}

%Next we show the existence of such Nijenhuis operators for $\Lie(\s)$.

We see that Equation \eqref{eq:SNijenhuis01} means that  $N_0$ is an
ordinary Nijenhuis operator for Lie algebra $\s$. The meanings of
the other two equations are given as follows.

\begin{pro}\label{pro:SNijenhuis} With the notations above, suppose that $N_0$ satisfies
\eqref{eq:SNijenhuis01} and $\langle\cdot,\cdot\rangle_\s$ is
invariant under  the operator $T_\lambda=\id+\lambda N_0$, i.e.
$\langle T_\lambda x,T_\lambda y\rangle_\s=\langle x,y\rangle_\s$,
where $\lambda \in \R$ is a parameter,  then $N=(N_0, 0)$ is a
Nijenhuis operator for the Lie $2$-algebra $\Lie(\s)$.
\end{pro}
\pf
Since $\langle\cdot,\cdot\rangle_\s$ is invariant under $T_\lambda$, we have
\begin{eqnarray}\label{pro-eq}
\langle N_0x,y\rangle_\s+\langle x,N_0y\rangle_\s=0, \quad \langle
N_0x,N_0y\rangle_\s=0,
\end{eqnarray}
which means that $ N_0$ is skew-symmetric with respect to
$\langle\cdot,\cdot\rangle_\s$, and $N'_0\circ N_0=0$. Thus, we have
$ N^2_0=0. $
Therefore, by \eqref{eq:SNijenhuis01}, we have
$$\langle[N_0x,N_0y]_\s,N_0z\rangle_\s=\langle N_0[x,y]_N, N_0z\rangle_\s=\langle N^2_0[x,y]_N, z\rangle_\s=0,$$
which implies that  \eqref{eq:SNijenhuis02} holds. Also by
\eqref{eq:SNijenhuis01}, we have
\begin{eqnarray*}
&&\langle[N_0x,N_0y]_\s,z\rangle_\s+\langle[N_0x,y]_\s,N_0z\rangle_\s+\langle[x,N_0y]_\s,N_0z\rangle_\s\\
&=&\langle[N_0x,N_0y]_\s,z\rangle_\s-\langle N_0[N_0x,y]_\s,z\rangle_\s-\langle N_0[x,N_0y]_\s,z\rangle_\s\\
&=&\langle[N_0x,N_0y]-N_0[N_0x,y]_\s- N_0[x,N_0y]_\s,z\rangle_\s\\
%&=&\langle[N_0x,N_0y]_\s- N_0[N_0x,y]_\s- N_0[x,N_0y]_\s,z\rangle_\s\\
&=&-\langle N^2_0[x,y]_\s,z\rangle_\s\\%=-\langle N_0[x,y]_\s,N_0z\rangle_\s\\
&=&0.
\end{eqnarray*}
Thus, \eqref{eq:SNijenhuis03} holds. The proof is finished. \qed

\begin{cor}\label{cor:Nij}
  With the notations above, suppose that $N_0$ satisfies
\eqref{eq:SNijenhuis01}. If $N_0$ is skew-symmetric and satisfies  $N_0^2=0$, then $N=(N_0, 0)$ is a
Nijenhuis operator for the Lie $2$-algebra $\Lie(\s)$.
\end{cor}

\begin{ex}{\rm
We now construct a class of  Nijenhuis operators for  Lie 2-algebra
$\Lie(\s)$, $\s=\h\oplus \h^*$, for  a finite dimensional Lie
algebra $\h$, where $\h^*$ is the dual space of $\h$ with zero Lie
bracket. It is called a Lie 2-algebra of string type in \cite{shengzhu3}. Recall that $\s=\h\oplus \h^*$ is a quadratic Lie algebra
defined by, for all $ x, y\in \h, ~ \xi,  \eta\in \h^*$,
$$[x+\xi, y+\eta]_\s=[x,y]+\ad^*_{x}\eta-\ad^*_{y}\xi, \quad \langle x+\xi, y+\eta\rangle_\s=\langle x, \eta\rangle+\langle\xi,y\rangle. $$
 Let $H = -H^*:
\h\to \h^*,$ be a skew-symmetric linear map. We use  $N_0$ to denote
its extension on  $\h\oplus \h^*$, i.e., $N_0(x+\xi)=Hx$, or in term of a matrix: $N_0=\left(\begin{array}{cc}
  0&0\\H&0
\end{array}\right)$.
We have
\begin{eqnarray*}
&&N_0([x+\xi,y+\eta]_N)\\
&=&N_0([N_0(x+\xi),y+\eta]_\s+[x+\xi,N_0(y+\eta)]_\s-N_0[x+\xi,y+\eta]_\s)\\
&=&N_0([Hx,y+\eta]+[x+\xi,Hy]-H[x,y])\\
&=&N_0(-\ad^*_{y}(Hx)+\ad^*_{x}(Hy)-H[x,y])\\
&=&0.
\end{eqnarray*}
On the other hand, $[N_0(x+\xi),N_0(y+\eta)]_\s=[Hx,Hy]=0$. Thus,
\eqref{eq:SNijenhuis01} holds, i.e., $N_0$ is an ordinary Nijenhuis
operator for Lie algebra $\s=\h\oplus\h^*$. Furthermore, it is obvious that $N_0$ is skew-symmetric and satisfies $N_0^2=0$, by Corollary \ref{cor:Nij},  $(N_0,0)$
is a Nijenhuis operator for the Lie 2-algebra $\Lie(\s)$. }
\end{ex}

 Now we construct another class of examples of Nijenhuis operators for $\Lie(\s)$ given in the above example in term of $\mathcal O$-operators. Let $\frkh$ be a Lie algebra, and $V$ be a vector space. Let $\rho:\frkh\longrightarrow\gl(V)$ be a representation. A linear operator $T:V\rightarrow \frkh$ is called an {\bf $\mathcal O$-operator} associated to $( V,\rho)$ if $T$ satisfies
\begin{equation}\label{eq:Ooperator}
  T(\rho(Tu)v+\rho(Tv)u)=[Tu,Tv]
\end{equation}
Such a concept was
introduced  to study the classical Yang-Baxter equations
and integrable systems. For more information, see \cite{O-operator}.
\begin{ex}{\rm
  Let $T = -T^*:
\h^*\to \h,$ be a skew-symmetric $\mathcal O$-operator associated to the coadjoint representation $(\frkh^*,\ad^*)$. We use  $N_0$ to denote
its extension on  $\h\oplus \h^*$, i.e., $N_0(x+\xi)=T\xi$, or in term of a matrix: $N_0=\left(\begin{array}{cc}
  0&T\\0&0
\end{array}\right)$.
it is straightforward to check that
\eqref{eq:SNijenhuis01} holds. Also by Corollary \ref{cor:Nij},  $(N_0,0)$
is a Nijenhuis operator for the Lie 2-algebra $\Lie(\s)$. }
\end{ex}

\begin{defi}
A deformation is said to be trivial if there exists  linear maps $N_0:\frkg_0\to \frkg_0,N_1: \frkg_{-1}\to \frkg_{-1}$,
and $N_2:\wedge^2 \frkg_0\to \frkg_{-1}$,
such that $(T_0,T_1,T_2)$ is a morphism from $(\g,\dM\ulam,[\cdot,\cdot]_\lambda, l^\lambda_3)$ to $(\g,\dM_\g,[\cdot,\cdot]_\g, l^\g_3) $, where $T_0 = \id + \lambda N_0$, $T_1 = \id + \lambda N_1$ and $T_2 = \lambda N_2$.
\end{defi}

Note that $(T_0,T_1,T_2)$ is a morphism means that
\begin{eqnarray}
\label{eq:trivialdeform0} \dM_\g \circ T_1(a)&=&T_0\circ \dM\ulam(a),\\
\label{eq:trivialdeform1} T_0[x,y]_\lambda&=&[T_0 x,T_0 y]_\g+\dM_\g T_2(x,y),\\
\label{eq:trivialdeform2} T_1[x,a]_\lambda&=&[T_0 x,T_1 a]_\g+T_2(x,\dM\ulam a),\\
\label{eq:trivialdeform3} T_2([x,y]_\lambda,z)+c.p.+T_1 l^\lambda_3(x,y,z)&=&[T_0(x),T_2(y,z)]_\g+c.p.+l_3^\g(T_0 x,T_0 y,T_0 z).
\end{eqnarray}
Now we consider conditions that $N=(N_0,N_1,N_2)$ should  satisfy.
For \eqref{eq:trivialdeform0}, we have
$$\dM_\g a+\lambda \dM_\g N_1(a)=\dM_\g a+\lambda N_0(\dM_\g a)+\lambda \omega_1(a)+\lambda^2N_0\omega_1(a).$$
Thus, we have
$$\omega_1a=\dM_\g N_1a-N_0\dM_\g a,$$
$$N_0\omega_1a=0.$$
It follows that $N$ must satisfy the following condition:
\begin{align}\label{Nijenhuis0}
N_0(\dM_\g N_1a-N_0\dM_\g a)=0.
\end{align}

For \eqref{eq:trivialdeform1}, the left hand side is equal to
$$[x,y]_\g+\lambda N_0([x,y]_\g)+\lambda\omega^0_2(x,y)+\lambda^2 N_0\omega^0_2(x, y),$$
and the right hand side is equal to \begin{eqnarray*}
[x,y]_\g+\lambda [N_0x, y]_\g+\lambda [x,N_0y]_\g+\lambda^2 [N_0x,N_0y]_\g+\lambda\dM_\g N_2(x,y).
\end{eqnarray*}
Thus, \eqref{eq:trivialdeform1} is equivalent to
$$\omega^0_2(x, y) = [N_0x, y]_\g + [x, N_0y]_\g - N_0[x, y]_\g+\dM_\g N_2(x,y),$$
$$N_0\omega^0_2(x, y) = [N_0x, N_0y]_\g. $$
It follows that $N$ must satisfy the following condition:
\begin{align}\label{Nijenhuis1}
[N_0x, N_0y]_\g - N_0[N_0x, y]_\g - N_0[x, N_0y]_\g + N_0^2[x,y]_\g-N_0\dM_\g N_2(x,y)=0.
\end{align}

For \eqref{eq:trivialdeform2},  the left hand side is equal to
$$[x,a]_\g+\lambda \omega^1_2(x,a) +\lambda N_1([x,a]_\g)+\lambda^2 N_1\omega^1_2(x,a),$$
and the right hand side is equal to
$$[x,a]_\g+\lambda [N_0(x), a]_\g+\lambda [x,N_1(a)]_\g+\lambda^2 [N_0(x),N_1(a)]_\g+\lambda  N_2(x,\dM_\g a)+\lambda^2N_2(x,\omega_1a).$$
Thus, \eqref{eq:trivialdeform2} is equivalent to
$$\omega^1_2(x, a) = [N_0x, a]_\g + [x, N_1a]_\g - N_1[x, a]_\g+ N_2(x,\dM_\g a),$$
$$N_1\omega^1_2(x, a) = [N_0x, N_1a]_\g+N_2(x,\omega_1a). $$
It follows that $N$ must satisfy the following condition:
\begin{align}\label{Nijenhuis2}
[N_0x, N_1a]_\g+N_2(x,\omega_1a) - N_1[N_0x, a]_\g - N_1[x, N_1a]_\g + N_1^2[x,a]_\g-N_1N_2(x,\dM_\g a)=0.
\end{align}

For \eqref{eq:trivialdeform3}, the left hand side is equal to
\begin{eqnarray*}
&&l_3^\g(x,y,z)+\lambda\omega_3(x,y,z)+ \lambda N_1l_3^\g(x,y,z)+\lambda^2 N_1\omega_3(x,y,z)\\
&&+\lambda N_2([x,y]_\g,z)+c.p.+\lambda^2 N_2(\omega^0_2(x,y),z)+c.p.
%&=&l_3(x,y,z)+\lambda\omega_3(x,y,z)+ \lambda N_1(l_3(x,y,z))+\lambda^2 N_1(\omega_3(x,y,z))\\
%&&+\lambda N_2([x,y],z)+c.p.+\lambda^2 N_2([N_0x, y] + [x, N_0y] - N_0[x, y]+d N_2(x,y),z)+c.p.
\end{eqnarray*}
and the right hand side is equal to
\begin{eqnarray*}
&&l_3^\g(x,y,z)+\lambda l_3^\g(N_0x,y,z)+c.p.+\lambda^2 l_3^\g(N_0x,N_0y,z)+c.p.\\
&&+\lambda^3l_3^\g(N_0x,N_0y,N_0z)+\lambda[x, N_2(y,z)]_\g+c.p.+\lambda^2[N_0x, N_2(y,z)]_\g+c.p.
\end{eqnarray*}
Thus,  \eqref{eq:trivialdeform3} is equivalent to
\begin{eqnarray*}
  \omega_3(x,y,z)=l_3^\g(N_0x,y,z)+c.p.-N_1l_3^\g(x,y,z)+[x, N_2(y,z)]_\g+c.p.- (N_2([x,y]_\g,z)+c.p.),\\
  N_1\omega_3(x,y,z)=l_3^\g(N_0x,N_0y,z)+c.p.+[N_0x, N_2(y,z)]_\g+c.p.-(N_2(\omega^0_2(x,y),z)+c.p.),\\
  l_3^\g(N_0x,N_0y,N_0z)=0.
\end{eqnarray*}
It follows that $N$ must satisfy the following conditions:
\begin{eqnarray}\label{Nijenhuis3}
\nonumber&N_1(l_3^\g(N_0x,y,z))+c.p.-N^2_1l_3^\g(x,y,z)+N_1([x,N_2(y,z)]_\g)+c.p.-(N_1N_2([x,y]_\g,z)+c.p.)\\\label{Nijenhuis3}
&-(l_3^\g(N_0x,N_0y,z)+c.p.)-([N_0x, N_2(y,z)]_\g+c.p.)+N_2(\omega^0_2(x,y),z)+c.p.=0,\\\label{Nijenhuis33}
& l_3^\g(N_0x,N_0y,N_0z)=0.
\end{eqnarray}

Thus,  $(T_0,T_1,T_2)$ is a morphism if and only if
  $N=(N_0,N_1,N_2)$ satisfy conditions \eqref{Nijenhuis0}, \eqref{Nijenhuis1}, \eqref{Nijenhuis2},  \eqref{Nijenhuis3} and \eqref{Nijenhuis33}.  Note that in this case, $N=(N_0,N_1,N_2)$ is not a Nijenhuis operator.

  \begin{rmk}
    A trivial deformation does not give rise to a Nijenhuis operator. This result is different from the case of ordinary Lie algebras. However, the converse is true, i.e. a Nijenhuis operator could give a trivial deformation.
  \end{rmk}

A Nijenhuis operator $(N_0,N_1)$ could give a trivial deformation by setting
\begin{equation}\label{eq:omega exact}
(\omega_1,\omega^0_2,\omega^1_2,\omega_3)=D(N_0,N_1).
\end{equation}
%$$\omega_2+\psi_2+\nu_2+\theta_2=D_{1}(\omega_1+\psi_1+\theta_1).$$
%In fact,
%\begin{eqnarray*}
%\omega_1&=&\widehat{d}N_0+\widehat{\partial}N_1\in \Hom(\frkg_{-1},V_{0}),\\
%\omega^0_2&=&dN_0+\widehat{d}N_2\in \Hom(\wedge^2\frkg_{0},V_{0}),\\
%\omega^1_2&=&dN_0+\widehat{d}N_2\in\Hom(\frkg_{0}\wedge\frkg_{-1},V_{-1}),\\
%\omega_3&=&d_{l_3}N_1+d_{l_2}N_2\in \Hom(\wedge^3\frkg_{0},V_{-1}).
%\end{eqnarray*}
\begin{thm}\label{thm:Nijenhuis}
Let $N=(N_0,N_1)$ be a Nijenhuis operator. Then a deformation
can be obtained by putting
\begin{equation}\label{Nijenhuis}
\left\{\begin{array}{rll}
\omega_1&=&0,\\
\omega^0_2(x, y)&=&[N_0x, y]_\g + [x, N_0y]_\g - N_0[x, y]_\g,\\
\omega^1_2(x, a)&=&[N_0x, a]_\g + [x, N_1a]_\g - N_1[x, a]_\g\\
\omega_3(x,y,z)&=&l_3^\g(N_0x,y,z)+c.p.-N_1(l_3^\g(x,y,z)).\\
\end{array}\right.
\end{equation}
Furthermore, this deformation is trivial.
\end{thm}
\pf Since $(\omega_1,\omega^0_2,\omega^1_2,\omega_3)=D(N_0,N_1)$, it is obvious that $(\omega_1,\omega^0_2,\omega^1_2,\omega_3)$ is closed. By Proposition \ref{pro:Nijenhuis}, $(\omega_1,\omega^0_2,\omega^1_2,\omega_3)=(0,\dM_N,l_3^N)$ defines a Lie 2-algebra structure. By Theorem \ref{thm:deformation}, $(\omega_1,\omega^0_2,\omega^1_2,\omega_3)$ generates a deformation.     \qed

\begin{ex}{\rm
  Let $(\g;\dM_\g=0,[\cdot,\cdot]_\g,l_3^\g)$ be a skeletal Lie 2-algebra. Then $(N_0,N_1)$ is a Nijenhuis operator if and only if the Nijenhuis conditions (ii)-(vi) hold.

  Note that $\omega_1=0$ in this case, which implies that a trivial deformation of a skeletal Lie 2-algebra is still skeletal. }
\end{ex}

\begin{ex}{\rm
  Let $(\g;\dM_\g,[\cdot,\cdot]_\g,l_3^\g=0)$ be a strict Lie 2-algebra. Then $(N_0,N_1)$ is a Nijenhuis  if and only if the Nijenhuis conditions (i)-(iii) hold. It is obvious that $l_3^N=0$ in this case, which implies that a trivial deformation given by a Nijenhuis operator is also strict.

 However, for a general trivial deformation, $\omega_3$ could be nonzero, which implies that a trivial deformation of a strict Lie 2-algebra maybe nonstrict. }
\end{ex}

\section{ Abelian extensions of Lie 2-algebras }

In this section, we study abelian extensions of Lie 2-algebras using the cohomology theory studied in Section \ref{sec:cohom}. Similar as the classical case, we show that associated to any abelian extension, there is a representation and a 2-cocycle. Consequently, any abelian extension can be viewed as a deformation of a semidirect product Lie 2-algebra, see Remark \ref{rmk:deform}.  Furthermore, abelian extensions can be classified by the second cohomology group.
For nonabelian extensions of Lie 2-algebras, see \cite{Sheng2}.

\begin{defi}
 \begin{itemize}
 \item[\rm (i)] Let $(\mathfrak{g},\dM_\g,[\cdot,\cdot]_\g,l_3^\g)$, $(\mathfrak{h},\dM_\frkh,[\cdot,\cdot]_\frkh,l_3^\frkh)$, $(\hat{\mathfrak{g}},\hat{\dM},[\cdot,\cdot]_{\hat{\g}},\hat{l_3})$ be Lie 2-algebras and
$i=(i_{0},i_{1}):\frkh\longrightarrow\hat{\mathfrak{g}},~~p=(p_{0},p_{1}):\hat{\mathfrak{g}}\longrightarrow\mathfrak{g}$
be strict homomorphisms. The following sequence of Lie 2-algebras is a
short exact sequence if $\mathrm{Im}(i)=\mathrm{Ker}(p)$,
$\mathrm{Ker}(i)=0$ and $\mathrm{Im}(p)=\g$.

\begin{equation}\label{eq:ext1}
\CD
  0 @>0>>  \frkh_{-1} @>i_1>> \hat{\g}_{-1} @>p_1>> \g_{-1} @>0>> 0 \\
  @V 0 VV @V \dM_\frkh VV @V \hat{\dM} VV @V\dM_\g VV @V0VV  \\
  0 @>0>> \frkh_{0} @>i_0>> \hat{\g}_0 @>p_0>> \g_0@>0>>0
\endCD
\end{equation}

We call $\hat{\mathfrak{g}}$  an extension of $\mathfrak{g}$ by
$\frkh$, and denote it by $\E_{\hat{\g}}.$
It is called an abelian extension if $\frkh$ is abelian, i.e. if $[\cdot,\cdot]_{\frkh}=0$ and $l^{\frkh}_3(\cdot,\cdot,\cdot)=0$. We will view $\frkh$ as subcomplex of $\g$ directly, and omit the map $i$.

\item[\rm (ii)] A splitting $\sigma:\mathfrak{g}\longrightarrow\hat{\mathfrak{g}}$ of $p:\hat{\mathfrak{g}}\longrightarrow\mathfrak{g}$
consists of linear maps
$\sigma_0:\mathfrak{g}_0\longrightarrow\hat{\g}_0$ and
$\sigma_1:\mathfrak{g}_{-1}\longrightarrow\hat{\g}_{-1}$
 such that  $p_0\circ\sigma_0=id_{\mathfrak{g}_0}$ and  $p_1\circ\sigma_1=id_{\mathfrak{g}_{-1}}$.
\item[\rm (iii)] Two extensions of Lie 2-algebras
 $\E_{\hat{\g}}:0\longrightarrow\frkh\stackrel{i}{\longrightarrow}\hat{\g}\stackrel{p}{\longrightarrow}\g\longrightarrow0$
 and $\E_{\tilde{\g}}:0\longrightarrow\frkh\stackrel{j}{\longrightarrow}\tilde{\g}\stackrel{q}{\longrightarrow}\g\longrightarrow0$ are equivalent,
 if there exists a Lie $2$-algebra homomorphism $F:\hat{\g}\longrightarrow\tilde{\g}$  such that $F\circ i=j$, $q\circ
 F=p$ and $F_2(i(u),\alpha)=0$, for all
 $u\in\frkh_0,~\alpha\in\hat{\g}_0$.
\end{itemize}
\end{defi}

Let $\hat{\mathfrak{g}}$ be an abelian extension of $\mathfrak{g}$ by
$\frkh$, and $\sigma:\g\longrightarrow\hat{\g}$ be a splitting.
Define $\mu_0,\mu_1,\mu_2$ by
\begin{equation}\label{eq:morphism}
\left\{\begin{array}{rlclcrcl}
\mu_{0}:&\mathfrak{g}_{0}&\longrightarrow& \End^{0}_{\dM_\frkh}(\frkh),&& \mu_{0}(x)(u+m)&\triangleq&[\sigma(x),u+m]_{\hg},\\
\mu_{1}:&\mathfrak{g}_{-1}&\longrightarrow& \End^{1}(\frkh),&&\mu_{1}(a)(u)&\triangleq&[\sigma(a),u]_{\hg},\\
\mu_{2}:&\wedge^2\mathfrak{g}_{0}&\longrightarrow&\End^{1}(\frkh),&& \mu_{2}(x,y)(u)&\triangleq&-\hat{l_{3}}(\sigma(x),\sigma(y),u),
\end{array}\right.
\end{equation}
for all $x,y\in\mathfrak{g}_{0}$, $a\in\mathfrak{g}_{-1}$,
$u\in\frkh_{0}$ and $m\in\frkh_{-1}$.

\begin{pro}\label{pro:2-modules}
With the above notations, $(\mu_0,\mu_1,\mu_2)$ is a homomorphism from $\g$ to the strict Lie $2$-algebra $\End(\h)$, i.e. $\g$ represents on $\h$ via $(\mu_0,\mu_1,\mu_2)$. Furthermore,  $(\mu_0,\mu_1,\mu_2)$ does not depend on the choice of the splitting $\sigma$.
Moreover,  equivalent abelian extensions give the same representation of $\g$ on $\h$.
\end{pro}

\pf
 By the fact that $p$ is a strict homomorphism, it is easy to show that $\mu_0,\mu_1,\mu_2$ are well-defined.

 Since $\h$ is abelian, we have
 \begin{eqnarray*}
   &&\mu_0([x,y]_\g)(u+m)-[\mu_0(x),\mu_0(y)](u+m)\\
   &=&[\sigma[x,y]_\g,u+m]_{\hg}-[\sigma(x),[\sigma(y),u+m]_{\hg}]_{\hg}+[\sigma(y),[\sigma(x),u+m]_{\hg}]_{\hg}\\
   &=&[[\sigma(x),\sigma(y)]_{\hg},u+m]_{\hg}-[\sigma(x),[\sigma(y),u+m]_{\hg}]_{\hg}+[\sigma(y),[\sigma(x),u+m]_{\hg}]_{\hg}\\
   &=&-\dM_{\frkh}\hat{l_3}(\sigma(x),\sigma(y),u)-\hat{l_3}(\sigma(x),\sigma(y),\dM_{\frkh}m),
 \end{eqnarray*}
which implies that
$$
\mu_0([x,y]_\g)-[\mu_0(x),\mu_0(y)]=\delta (\mu_2(x,y)).
$$
Similarly, we have
$$
\mu_1([x,a]_\g)-[\mu_0(x),\mu_1(a)]=\mu_2(x,\dM_\g a).
$$
By the Jacobiator identity, we can also show that
$$\mu_2([x,y]_\g,z)+c.p.+\mu_1(l_3^\g(x,y,z))=[\mu_0(x),\mu_2(y,z)]+c.p..$$
Thus, $(\mu_0,\mu_1,\mu_2)$ is a homomorphism.
\emptycomment{
First, since $i, p$ are strict morphisms, we have that $\h_1\cong\Ker p_1$, $\h_0\cong\Ker p_0$,
$p_0[\hat{x},\hat{y}]_{\hg}=[p_0\hat{x},p_0\hat{y}]_{\g}$, $p_1[\hat{x},\hat{a}]_{\hg}=[p_0\hat{x},p_1\hat{a}]_{\g}$, and $p_1\hat{l_{3}}(\hat{x},\hat{y},\hat{z})=l_3(p_0\hat{x},p_0\hat{y},p_0\hat{z})$.
Then, for $u\in\h_0$, $p_0u=0$, so $p_0[\sigma(x),u]_{\hg}=[p_0\sigma(x),p_0u]_{\hg}=0$, $[\sigma(x),u]_{\hg}\in  \ker p_0 \cong \h_0$.
Similarly, for $m\in\h_1$, $p_1m=0$, $[\sigma(x),m]_{\hg}\in  \ker p_1 \cong \h_1$. One can also check that $p_1[\sigma(a),u]_{\hg}=[p_1\sigma(a),p_0u]=0$, $p_1\hat{l_{3}}(\sigma(x),\sigma(y),u)=l_3(p_0\sigma(x),p_0\sigma(y),p_0u)=0$,
thus $[\sigma(a),u]_{\hg}\in \h_1,  \hat{l_{3}}(\sigma(x),\sigma(y),u)\in \h_1$.}

Since $\h$ is abelian, we can show that $\mu_{i}$ are independent of
the choice of $\sigma$. In fact, if we choose another splitting
$\sigma':\g\to\hg$, then $p_0(\sigma(x)-\sigma'(x))=x-x=0$,
$p_1(\sigma(a)-\sigma'(a))=a-a=0$, i.e. $\sigma(x)-\sigma'(x)\in
\Ker p_0=\h_0$, $\sigma(a)-\sigma'(a)\in \Ker p_1=\h_{-1}$. Thus,
$[\sigma(x)-\sigma'(x), u+m]_{\hat{\g}}=0$, $[\sigma(a)-\sigma'(a),
u]_{\hat{\g}}=0$, which implies that $\mu_0,\mu_1$ are independent
on the choice of $\sigma$. We also have $\hat{l_{3}}(\cdot,u,v)=0$
for all $u,v\in\frkh_0$, which implies that
$\hat{l_{3}}(\sigma'(x),\sigma'(y),u)=\hat{l_{3}}(\sigma(x),\sigma(y),u)$,
i.e. $\mu_2$ is also independent on the choice of $\sigma$.

Suppose that $\E_{\hat{\g}}$ and $\E_{\tilde{\g}}$ are equivalent abelian extensions, and $F:\hat{\g}\longrightarrow\tilde{\g}$ is the Lie 2-algebra homomorphism satisfying $F\circ i=j$, $q\circ
 F=p$ and $F_2(i(u),\alpha)=0$, for all  $u\in\frkh_0,~\alpha\in\hat{\g}_0$.
Choose linear sections $\sigma$ and $\sigma'$ of $p$ and $q$. Then we have $q_0F_0\sigma(x)=p_0\sigma(x)=x=q_0\sigma'(x)$,
so $F_0\sigma(x)-\sigma'(x)\in \Ker q_0=\h_0$. Thus, we have
$$[\sigma'(x),u+m]_{\tg}=[F_0\sigma(x),u+m]_{\tg}=F_0[\sigma(x),u+m]_{\hg}=[\sigma(x),u+m]_{\hg},$$
which implies that equivalent abelian extensions give the same $\mu_0$. Similarly, we can show that equivalent abelian extensions also give the same $\mu_1$. At last, by the fact that $F=(F_0,F_1,F_2)$ is a homomorphism and $F_2(u,\cdot)=0$, we have
$$
\hat{l_3}(\sigma(x),\sigma(y),u)=\widetilde{l_3}(F_0\sigma(x),F_0\sigma(y),u)=\widetilde{l_3}(\sigma'(x),\sigma'(y),u).
$$
Therefore, equivalent abelian extensions also give the same $\mu_2$. The proof is finished.
\qed\vspace{3mm}

Let $\sigma:\frkg\longrightarrow\hat{\frkg}$  be a
splitting of the abelian extension \eqref{eq:ext1}. Define the following linear maps:
$$
\begin{array}{rlclcrcl}
\psi:&\frkg_{-1}&\longrightarrow&\h_{0},&& \psi(a)&\triangleq&\hat{\dM}\sigma(a)-\sigma(\dM_\frkg a),\\
\omega:&\wedge^2\frkg_{0}&\longrightarrow&\h_{0},&& \omega(x,y)&\triangleq&[\sigma(x),\sigma(y)]_{\hat{\frkg}}-\sigma[x,y]_\frkg,\\
\nu:&\frkg_{0}\wedge\frkg_{-1}&\longrightarrow&\h_{-1},&& \nu(x,a)&\triangleq&[\sigma(x),\sigma(a)]_{\hat{\frkg}}-\sigma[x,a]_\frkg,\\
\theta:&\wedge^3\frkg_{0}&\longrightarrow&\h_{-1},&&
\theta(x,y,z)&\triangleq&\hat{l_{3}}(\sigma(x),\sigma(y),\sigma(z))-\sigma(l_{3}^\frkg(x,y,z)),
\end{array}
$$
for all $x,y,z\in\frkg_{0}$, $a\in\frkg_{-1}$, $u\in \h_{0}$ and
$m\in \h_{-1}$.

\begin{thm}\label{thm:2-cocylce}
Let $\E_{\hg}$ be an abelian extension of $\g$ by $\h$ given by \eqref{eq:ext1}, then $(\psi,\omega,\nu,\theta)$ is a $2$-cocycle of $\g$ with coefficients in $\frkh$, where the representation is given by $(\mu_0,\mu_1,\mu_2)$.
%Moreover, equivalent abelian extension give equivalent 2-cocycles.
\end{thm}

\pf
By the equality $\hat{\dM}[\sigma x,\sigma
a]_{\hat{\mathfrak{g}}}=[\sigma x,\hat{\dM}\sigma
a]_{\hat{\mathfrak{g}}}$, we obtain that
\begin{eqnarray}\label{eq:c1}
\mu_0(x)(\psi(a))+\omega(x,\dM_\mathfrak{g}a)&=&\psi([x,a]_\mathfrak{g})+\dM_\frkh\nu(x,a).
\end{eqnarray}

By the equality $[\hat{\dM}(\sigma a),\sigma
b]_{\hat{\mathfrak{g}}}=[\sigma a,\hat{\dM}\sigma
b]_{\hat{\mathfrak{g}}}$, we obtain that
\begin{eqnarray}\label{eqn:c2}
-\mu_1(b)(\psi(a))-\nu(b,\dM_\g a)=\mu_1(a)(\psi(b))+\nu(a,\dM_\g b).
\end{eqnarray}

By the equality
$$
[\sigma x, [\sigma y,\sigma
z]_{\hat{\mathfrak{g}}}]_{\hat{\mathfrak{g}}}+c.p.=\hat{\dM}\hat{l_{3}}(\sigma
x,\sigma y,\sigma z),
$$
we get
\begin{equation}\label{eq:c3}
\mu_{0}(x)\omega(y,z)+\omega(x,[y,z]_\mathfrak{g})+c.p.=\dM_\frkh
\theta(x,y,z)+\psi(l_3^\g(x,y,z)).
\end{equation}

We also have the equality
\begin{eqnarray*}
\label{eqn:xya}&&[\sigma x, [\sigma y,\sigma a]_{\hat{\mathfrak{g}}}]_{\hat{\mathfrak{g}}}+c.p.
=\hat{l_{3}}(\sigma x,\sigma y,\hat{\dM}\sigma a).
\end{eqnarray*}
Consider the left hand side, we have
\begin{eqnarray*}
&&[\sigma x, [\sigma y,\sigma a]_{\hat{\mathfrak{g}}}]_{\hat{\mathfrak{g}}}+c.p.\\
&=&[\sigma x,\sigma[y,a]_\g+\nu(y,a)]_{\hat{\g}}+[\sigma y,\sigma[a,x]_\g+\nu(a,x),]_{\hat{\g}}+[\sigma a,\sigma[x,y]_\g+\omega(x,y)]_{\hat{\g}}\\
&=&\sigma[x,[y,a]_\g]_\g+\nu(x,[y,a]_\g)+\mu_0(x)\nu(y,a)+\sigma[y,[a,x]_\g]_\g+\nu(y,[a,x]_\g)+\mu_0(y)\nu(a,x)\\
&&+\sigma[a,[x,y]_\g]_\g+\nu(a,[x,y]_\g)+\mu_1(a)\omega(x,y)\\
&=&\sigma l_3^\g(x,y,\dM_\g a)+\mu_0(x)\nu(y,a)+\mu_0(y)\nu(a,x)+\mu_1(a)\omega(x,y)\\
&&+\nu(x,[y,a]_\g)+\nu(y,[a,x]_\g)+\nu(a,[x,y]_\g).
\end{eqnarray*}
Consider the right hand side, we have
\begin{eqnarray*}
\hat{l_{3}}(\sigma x,\sigma y,\hat{\dM}\sigma a)
&=&\hat{l_{3}}(\sigma x,\sigma y,\sigma(\dM_\g a)+\psi(a))\\
&=&\sigma l_3^\g(x,y,\dM_\g a)+\theta(x,y,\dM_\g a)-\mu_2(x,y)\psi(a).
\end{eqnarray*}
Thus, we have
\begin{eqnarray}
\nonumber&&\mu_0(x)\nu(y,a)+\mu_0(y)\nu(a,x)+\mu_1(a)\omega(x,y)
+\nu(x,[y,a]_\g)+\nu(y,[a,x]_\g)+\nu(a,[x,y]_\g)\\\label{eq:c4}
&&-\theta(x,y,\dM_\g a)+\mu_2(x,y)\psi(a)=0.
\end{eqnarray}

At last, by the Jacobiator identity:
$$
\hat{l_{3}}([\sigma x,\sigma y]_{\hat{\g}},\sigma z,\sigma
t)+c.p.=[\sigma x,\hat{l_{3}}(\sigma y,\sigma z,\sigma
t)]_{\hat{\mathfrak{g}}}+c.p.,
$$
we obtain
\emptycomment
{
\begin{eqnarray*}
&&\hat{l_3}(\sigma[x,y]_\g-\omega(x,y),\sigma z,\sigma t)+c.p.\\
&=&\sigma
l_3^\g([x,y]_\g,z,t)-\theta([x,y]_\g,z,t)-\mu_2(z,t)\omega(x,y)+c.p.,
\end{eqnarray*}
and the right hand side is equal to
\begin{eqnarray*}
&&[\sigma x,\sigma l_3^\g(y,z,t)-\theta(y,z,t)]_{\hat{\g}}+c.p.\\
&=&\sigma[x,l_3^\g(y,z,t)]_\g-\nu(x,l_3^\g(y,z,t))-\mu_0(x)\theta(y,z,t)+c.p..
\end{eqnarray*}
}
\begin{eqnarray}\label{eq:c5}
\theta([x,y]_\g,z,t)-\mu_2(z,t)\omega(x,y)+c.p.=\mu_0(x)\theta(y,z,t)+\nu(x,l_3^\g(y,z,t))+c.p..
\end{eqnarray}
By \eqref{eq:c1}-\eqref{eq:c5}, we deduce that
$(\psi,\omega,\nu,\theta)$ is a 2-cocycle.\qed\vspace{3mm}

%\begin{lem}
%Let $\psi+\omega+\nu+\theta$ be a 2-cocycle  in the generalized Chevalley-Eilenberg complex of
%$\g$, Then we have an abelian extension of $\g$ by $V$ if where $\hg=\g\oplus V$ as direct sum of 2-term complex of
%vector spaces with Lie 2-algebra structures given by \eqref{eq:bracket}.
%Furthermore, equivalent 2-cocycles gives equivalent abelian extensions.
%\end{lem}

Now we can transfer the Lie 2-algebra
structure $(\hat{\dM},[\cdot,\cdot]_{\hat{\g}},\hat{l_3})$ on $\hat{\g}$ to the Lie 2-algebra structure $(\dM_{\gh},[\cdot,\cdot]_{\gh},l_3^{\gh})$ on $\g\oplus\frkh$ using the 2-cocycle given above. More precisely, we have
\begin{equation}\label{eq:bracket}
\left\{\begin{array}{rcl}
\dM_{\gh}(a+m)&\triangleq&\dM_\g(a)+\psi(a)+\dM_\frkh(m),\\
\brh{x+u,y+v}&\triangleq&[x,y]_\g+\omega(x,y)+\mu_{0}(x)v-\mu_{0}(y)u,\\
\brh{x+u,a+m}&\triangleq&[x,a]_\g+\nu(x,a)+\mu_{0}(x)m-\mu_{1}(a)u,\\
l_3^{\gh}(x+u,y+v,z+w)&\triangleq& l_3^\g(x,y,z)+\theta(x,y,z)
-\mu_{2}(x,y)(w)-\mu_{2}(z,x)(v)-\mu_{2}(y,z)(u),
\end{array}\right.
\end{equation}
for all $x,y,z\in\mathfrak{g}_{0}$, $a\in\mathfrak{g}_{-1}$,
$u,v,w\in\frkh_{0}$ and $m\in\frkh_{-1}$. Thus any extension
$E_{\hat{\g}}$ given by \eqref{eq:ext1} is isomorphic to
\begin{equation}\label{eq:ext2}
\CD
  0 @>0>>  \frkh_{-1} @>i_1>> \g_{-1}\oplus \frkh_{-1} @>p_1>> \g_{-1} @>0>> 0 \\
  @V 0 VV @V \dM_\frkh VV @V \hat{\dM} VV @V\dM_\g VV @V0VV  \\
  0 @>0>> \frkh_{0} @>i_0>> \g_0\oplus \frkh_0 @>p_0>> \g_0@>0>>0,
\endCD
\end{equation}
where the Lie 2-algebra structure on $\frkg\oplus\frkh$ is given by
\eqref{eq:bracket} for some morphism $(\mu_0,\mu_1,\mu_2)$ given by \eqref{eq:morphism},
$(i_0,i_1)$ is the inclusion and $(p_0,p_1)$ is the projection. We
denote the extension \eqref{eq:ext2} by
$\E_{\frkg\oplus\frkh}$.

\begin{rmk}\label{rmk:deform}
  In fact, the extension $\E_{\frkg\oplus\frkh}$ can be viewed as a deformation of the semidirect product Lie 2-algebra $\g\ltimes_\mu \h$ by the 2-cocycle $(\psi,\omega,\nu,\theta)$.  First, given a representation $\mu$ of $\g$ on $\h$, we can obtain the semidirect product Lie 2-algebra $\g\ltimes_\mu \h$:
  \begin{equation}\label{eq:bracketsimi}
\left\{\begin{array}{rcl}
\dM_s(a+m)&\triangleq&\dM_\g(a)+\dM_\frkh(m),\\
~[x+u,y+v]_s&\triangleq&[x,y]_\g+\mu_{0}(x)v-\mu_{0}(y)u,\\
~[x+u,a+m]_s&\triangleq&[x,a]_\g+\mu_{0}(x)m-\mu_{1}(a)u,\\
l_3^s(x+u,y+v,z+w)&\triangleq& l_3^\g(x,y,z)
-\mu_{2}(x,y)(w)-\mu_{2}(z,x)(v)-\mu_{2}(y,z)(u).
\end{array}\right.
\end{equation}
Then it is not hard to show that the $2$-cocycle  $(\psi,\omega,\nu,\theta)$ can be extended to the 2-cocycle $(\overline{\psi},\overline{\omega},\overline{\nu},\overline{\theta})$  of $\g\ltimes_\mu \h$ with the coefficients in the adjoint representation:
\begin{eqnarray*}
  \overline{\psi}(a+m)=\psi(a),\quad  \overline{\omega}(x+u,y+v)=\omega(x,y), \\
  \overline{\nu}(x+u,a+m)=\nu(x,a),\quad \overline{\theta}(x+u,y+v,z+w)=\theta(x,y,z).
\end{eqnarray*}
Now it is straightforward to see that the Lie 2-algebra structure \eqref{eq:bracket} is the deformation of $\g\ltimes_\mu \h$ by the 2-cocycle  $(\overline{\psi},\overline{\omega},\overline{\nu},\overline{\theta})$.
\end{rmk}

In the sequel, we only consider the abelian extensions in the form of \eqref{eq:ext2}, i.e. $(i_0,i_1)$ is the inclusion and $(p_0,p_1)$ is the projection. We fix the representation $(\mu_0,\mu_1,\mu_2)$ and study the relation between equivalent classes of abelian extensions and the second cohomology group $\mathbf{H}^2(\frkg;\mu)$.

\begin{thm}
Given a representation $(\mu_0,\mu_1,\mu_2):\g\longrightarrow \End(\h)$, then there is a one-to-one correspondence between equivalence classes of abelian extensions, in the form of \eqref{eq:ext2}, of the Lie $2$-algebra $\frkg$ by $\h$ and the second cohomology group $\mathbf{H}^2(\frkg;\mu)$.
\end{thm}

\pf
Let $\E'_{\gh}$ be another abelian extension determined by the 2-cocycle $(\psi',\omega',\nu',\theta')$. Denote the corresponding Lie 2-algebra structure on $\g\oplus \h$ by $(\dM',[\cdot,\cdot]',l_3')$. We only need to show that $\E_{\gh}$ and $\E'_{\gh}$ are equivalent if and only if 2-cocycles  $(\psi,\omega,\nu,\theta)$ and $(\psi',\omega',\nu',\theta')$ are in the same cohomology class.

If $\E_{\gh}$ and $\E'_{\gh}$ are equivalent, let
$F=(F_0,F_1,F_2):\E_{\gh}\longrightarrow\E'_{\gh}$
be the corresponding homomorphism.

 Since $F$ is an equivalence of extensions, we have
$$F_2(u,v)=0,\quad F_2(x,u)=0,\quad F_2(x,y)\in\frkh_{-1},$$
and there exist two linear maps $b_0:\g_0\longrightarrow\frkh_0$
and $b_1:\g_{-1}\longrightarrow\frkh_{-1}$ such that
 $$F_0(x+u)=x+b_0(x)+u,\quad
F_1(a+m)=a+b_1(a)+m.$$ Set  $b_2=F_2|_{\wedge^2\g_0}$.

First, by the equality
\begin{eqnarray*}
\label{eqn:fi} \dM'F_1(a)&=&F_0\dM_{\gh}(a),
%\label{eqn:F0F1} F_1(\brh{u,a})-\brt{F_0(u),F_1(a)}&=&0,
\end{eqnarray*}
we have
\begin{equation}\label{eq:exact1}
\psi(a)-\psi'(a)=\dM_\frkh b_1(a)-b_0(\dM_\g a).
%\quad\mu_1(a)-\mu_1'(a)&=&\ad_{\psi_1(a)}.
\end{equation}

 Furthermore, we have
$$
F_0\brh{x,y}-\brt{F_0(x),F_0(y)}=\dM'F_2(x,y),
$$
which implies that
\begin{equation}\label{eq:exact2}
\omega(x,y)-\omega'(x,y)
=\mu_0(x)b_0(y)-\mu_0(y)b_0(x)-b_0[x,y]_\g+\dM_\frkh\circ
b_2(x,y).
\end{equation}

Similarly, by $F_1\brh{x,a}-\brt{F_0(x),F_1(a)}=F_2(x,\hat{\dM}a)$,
we get
\begin{equation}\label{eq:exact3}
\nu(x,a)-\nu'(x,a)=\mu_0(x)b_1(a)-\mu_1(a)b_0(x)-b_1[x,a]_\g+b_2(x,\dM_\g a).
\end{equation}

At last, by the equality
$$\brt{F_0(x),F_2(y,z)}+c.p.+l_3'(F_0(x),F_0(y),F_0(z))=F_2(\brh{x,y},z)+c.p.+F_1l_3^{\gh}(x,y,z),$$
 we have
\begin{equation}\label{eq:exact4}
(\theta-\theta')(x,y,z)=\mu_0(x)b_2(y,z)-b_2([x,y]_\g,z)-\mu_2(x,y)b_0(z)+c.p.-b_1(l_3^\g(x,y,z)).
\end{equation}

By \eqref{eq:exact1}-\eqref{eq:exact4}, we deduce that $(\psi,\omega,\nu,\theta)-(\psi',\omega',\nu',\theta')=D(b_0,b_1,b_2)$. Thus, they are in the same cohomology class.

Conversely, if  $(\psi,\omega,\nu,\theta)$ and $ (\psi',\omega',\nu',\theta')$ are in the same cohomology class, assume that $(\psi,\omega,\nu,\theta)-(\psi',\omega',\nu',\theta')=D(b_0,b_1,b_2)$. Then define $(F_0,F_1,F_2)$ by
$$F_0(x+u)=x+b_0(x)+u,\quad
F_1(a+m)=a+b_1(a)+m,\quad F_2(x+u,y+v)=b_2(x,y). $$
Similar as the above proof, we can show that $(F_0,F_1,F_2)$ is an equivalence. We omit the details.
\qed

\end{document}